\documentclass[
aip, 
cha,
 amsmath,amssymb,
 reprint,%
]{revtex4-1}

\usepackage{graphicx}
\usepackage{dcolumn}
\usepackage{bm}

\usepackage[utf8]{inputenc}
\usepackage[T1]{fontenc}
\usepackage{mathptmx}
\usepackage{etoolbox}
\usepackage[printonlyused, nohyperlinks]{acronym}
\usepackage{mathtools}
\usepackage[hidelinks, linkcolor=blue, colorlinks=true, citecolor=blue, urlcolor=blue, pdfauthor={Yannick Schöhs}]{hyperref}
\usepackage{caption}
\usepackage{subcaption}
\usepackage{ragged2e}
\usepackage{float}
\usepackage[percent]{overpic}

\makeatletter
\def\@email#1#2{%
 \endgroup
 \patchcmd{\titleblock@produce}
  {\frontmatter@RRAPformat}
  {\frontmatter@RRAPformat{\produce@RRAP{*#1\href{mailto:#2}{#2}}}\frontmatter@RRAPformat}
  {}{}
}%
\makeatother
\begin{document}


\title{Origin of Frequency Clusters and Self-Organized Triplet Locking in the Kuramoto Model with Inertia}
\author{Yannick Schöhs}
  \email{yannick.schoehs@tum.de}

\author{Nicolas Thomé}%

\author{Katharina Krischer}
\noaffiliation
\affiliation{ 
Technische Universität München, School of Natural Sciences, Nonequilibrium Chemical Physics, James-Franck-Str. 1, D-85748 Garching, Germany
}%
 \email{krischer@tum.de}


\date{\today}

\begin{abstract}
   We investigate the origin of frequency clusters – states in which multiple groups of oscillators with different mean frequencies coexist – in the globally coupled Kuramoto model with inertia and identical oscillators. Two frequency clusters are studied in the thermodynamic limit, three frequency clusters with a system of seven oscillators. Using bifurcation analysis, we demonstrate that in both cases the frequency clusters emerge through homoclinic bifurcations. In the case of three frequency clusters, this necessarily entails the formation of a triplet locked state, characterized by rational relations among the mean frequency differences. The individual clusters may lose phase-synchrony via either transcritical or period-doubling bifurcations. 
    Finally, we establish that Hopf bifurcations cannot generate frequency clusters in phase oscillator systems. Instead, they can only arise through global bifurcations.

\end{abstract}

\maketitle


\section{Introduction}

Synchronization is one of the most fascinating phenomena of self-organization. It describes the spontaneous adjustment of rhythms that results in coherent, collective behavior among interacting dynamical units. 
Synchronization transforms microscopic variability into macroscopic order without centralized control \cite{Pikovsky.2010}. 
In networks of oscillators, intriguing phenomena can also occur when the oscillators partially lose synchrony. Systems of identical oscillators can form phase clusters or chimera states, for example. The former consist of internally synchronized groups of oscillators that have a constant phase difference to the oscillators in the other groups~\cite{Okuda.1993}. Chimeras are characterized by the coexistence of a synchronized and an incoherent oscillator group~\cite{Kuramoto.2002, Abrams.2004}. While we now have a fairly good understanding of how these states emerge and what properties they exhibit, much less is known about another type of synchronization pattern: so-called frequency clusters, which consist of frequency-synchronized groups, each with a different mean frequency.
Frequency clusters occur in a variety of systems,  among them chemical system~\cite{Patzauer.2021}, mechanical oscillators~\cite{Ebrahimzadeh.2020}, and  perhaps  most impressively and importantly, the human brain~\cite{Pfurtscheller.1999, Siegel.2012}.
In theoretical studies, frequency clusters have been found, for example, in the Wilson-Cowan model describing neuronal populations~\cite{Singh.2016}, in identical Stuart-Landau oscillators~\cite{Premalatha.2015} or in adaptively coupled Kuramoto-Sakaguchi oscillators~\cite{Berner.2019}. 

Unlike other cluster types, the formation of frequency clusters requires that the effective coupling strength between the oscillators depends on the current state of the oscillators, rendering the coupling adaptive. The arguably easiest oscillator network that supports frequency clusters is a network of identical Kuramoto oscillators with inertia~\cite{Berner.2021}. The \ac{kmi}~\cite{Tanaka.1997, Tanaka.1997b} is an extension of the well-studied Kuramoto model~\cite{Kuramoto.1975, Haken.1984, Strogatz.2000}, obtained by adding a second-order term that allows for frequency adaptation~\cite{Ermentrout.1991}.

Recently, the emergence of two frequency clusters in networks of identical Kuramoto oscillators with inertia was investigated in small, finite systems \cite{Maistrenko.2017,Brezetsky.2021,Ashwin.2025}, where these states are classified as (weak) chimeras. We perceive the in-phase or weak chimeras in small systems as a specific case of frequency clusters, as well as recently studied solitary states~\cite{Munyayev.2022}, in which only one oscillator exhibits a different frequency compared to a large synchronized group. 
In a system with three oscillators, Ashwin and Bick~\cite{Ashwin.2025} performed a bifurcation analysis and showed that in this system, frequency clusters arise from homoclinic bifurcations, akin to their formation in a two-cluster population network, as demonstrated by Belykh et al.~\cite{Belykh.2016}.
Furthermore, the transversal stability of solutions with two phase-synchronous clusters has been studied in the thermodynamic limit using evaporation eigenvectors \cite{Munyayev.2024}.  

In this paper, we provide an extended analysis of the origin of frequency clusters in a system of globally coupled identical Kuramoto oscillators with inertia, first extending results on two frequency clusters, then elucidating the origin of three frequency clusters. Building upon the above mentioned works \cite{Munyayev.2024, Ashwin.2025}, we perform a bifurcation analysis of two frequency clusters in cluster subspaces, thereby providing a stability analysis in the thermodynamic limit. 
Then, we turn the analysis to three frequency clusters. We elaborate that three frequency clusters form triplet-locked solutions and compare them to the well-studied Arnol'd tongues. On the other hand, we show how two and three frequency clusters lose their phase synchrony in both transversal period-doubling and transcritical bifurcations. Last, we argue that global bifurcations are required for the formation of frequency clusters.

Our work is structured as follows: In section~\ref{sec:Preliminaries}, we introduce the
theoretical and numerical frameworks and relevant terminology. In section~\ref{sec:2-freq}, we perform a bifurcation analysis of the 2-cluster state and study how two frequency clusters lose their internal phase-synchrony. Section~\ref{sec:3_frequency_clusters} deals with the longitudinal and transversal stabilities of three frequency clusters, which is again studied employing bifurcation analysis. In section~\ref{sec:Impossibility_of_a_Hopf_Bifurcation_to_Create_Frequency_Clusters}, we show that the formation of frequency clusters necessarily involves global bifurcations. Section~\ref{sec:C&O} summarizes our results and discusses future directions.


\section{Preliminaries \label{sec:Preliminaries}}
\subsection{Kuramoto Model with Inertia}
\label{sec:KMI}
We investigate a network of $N$ globally coupled, identical phase oscillators, which is described by the \ac{kmi} \cite{Tanaka.1997,Maistrenko.2017}. Here, the dynamics of the phase $\phi_i$ of oscillator $i$ is determined by the equation
\begin{equation}
	\label{eq:KMI_with_all_parameters}
	M \ddot{\phi}_i + \epsilon \dot{\phi}_i = \epsilon \omega - \frac{\sigma}{N} \sum^N_{j=1} \sin(\phi_i - \phi_j + \beta),
\end{equation}
where $M$ is the inertia coefficient, $\epsilon$ the damping constant, $\omega$ the intrinsic frequency, which is identical for all oscillators, $\sigma$ the coupling strength and $\beta$ the phase lag parameter which delays the interaction. 
To simplify the equations, we switch to the rotating reference frame with the transformation $\phi_i \rightarrow \phi_i + \omega \cdot t$, thereby eliminating the term $\epsilon \cdot \omega$.
The inertia coefficient $M$ is fixed at the value $M=1$ for the rest of this work, which corresponds to a rescaling of the parameters $\epsilon$ and $\sigma$ \cite{Ashwin.2025}. Thus, the influence of inertia is absorbed in  $\epsilon$ and $\sigma$. 
In this study, both of them are set to the small value $\epsilon=\sigma=0.05$, giving the inertia term a significant influence. 

Equations \eqref{eq:KMI_with_all_parameters} have several symmetries: (a) They are invariant under the transformations $\beta \rightarrow \beta + 2\pi$, and $\beta \rightarrow -\beta,\ \phi_i \rightarrow -\phi_i$. Therefore, $\beta$ can be restricted to the interval $\beta \in [0,\pi]$. 
(b) Since we consider identical oscillators, the equations are equivariant under the action of the permutation group $\mathbb{S}_N$, i.e. the system remains unchanged under the permutation of oscillators \cite{Golubitsky.2002}. 
(c) They are invariant under the discrete phase shift of a single oscillator by $2\pi$; $\phi_i \rightarrow \phi_i + 2\pi$. Therefore, we can restrict the phase variables to the circle $S^1$, such that $\phi_i \in [0, 2\pi)$. The system also exhibits a continuous phase-shift symmetry and is thus equivariant under a phase shift of all oscillators $\boldsymbol{\phi} \rightarrow \boldsymbol{\phi} + \theta \cdot \mathbf{1}_N$, where $\boldsymbol{\phi}$ is the vector containing the phases of all $N$ oscillators, $\theta \in \mathbb{R}$ and $\mathbf{1}_N$ is the $N$-dimensional vector whose entries are all equal to 1.

\subsubsection{Phase-Difference Coordinates}
Since equations~\eqref{eq:KMI_with_all_parameters} obey phase shift symmetry, they only depend on the phase differences. 
Thus, phase-difference coordinates can be introduced. Before changing coordinates, equations~\eqref{eq:KMI_with_all_parameters} are first reduced to a set of $2N$ coupled first-order \acp{ode}. As elaborated in the previous section, we set $M=1$ and $\omega=0$: 
\begin{subequations}
\label{eq:KMI_final_first_order_reduction}
\begin{equation}
	\label{eq:KMI_final_reduced_first_order_equations1_phase}
	\dot{\phi}_i = \psi_i    
\end{equation}  
\begin{equation}
	\label{eq:KMI_final_reduced_first_order_equations2_phase_velocity}
	\dot{\psi}_i = -\epsilon \psi_i - \frac{\sigma}{N} \sum^N_{j=1} \sin(\phi_i - \phi_j + \beta),
\end{equation}    
\end{subequations}
where $\psi_i$ is the phase velocity or frequency of oscillator $i$. The phase-difference coordinates are here defined with respect to the first oscillator $\Delta \phi_{i}\coloneqq\phi_1 - \phi_{i+1}, i \in \{1,\ldots,N-1\}$. Similarly, the phase velocity differences, or frequency differences are written as $\Delta \psi_{i}\coloneqq\psi_1 - \psi_{i+1}, i \in \{1,\ldots,N-1\}$. The \acp{ode} for the phase differences can be derived by subtracting equation \eqref{eq:KMI_final_reduced_first_order_equations1_phase} resp. \eqref{eq:KMI_final_reduced_first_order_equations2_phase_velocity} for oscillator $i$ from the equation \eqref{eq:KMI_final_reduced_first_order_equations1_phase} resp. \eqref{eq:KMI_final_reduced_first_order_equations2_phase_velocity} for oscillator 1, and then applying the definitions of the phase-difference coordinates. This yields the set of equations
\begin{subequations}
\label{eq:KMI_final_phase_difference_coordinates}
\begin{equation}
	\label{eq:KMI_final_phase_difference_coordinates_phase_difference}
    \Delta\dot{\phi}_i = \Delta\psi_i
\end{equation}
\begin{eqnarray}
    \Delta\dot{\psi}_i = &&-\epsilon \Delta\psi_{i} + \frac{\sigma}{N} \bigg[ \sin(-\Delta \phi_{i}+\beta) - \sin(\beta) \nonumber \\
     &&+ \sum^{N-1}_{j=1} \sin(\Delta\phi_{j} - \Delta\phi_{i} + \beta) -\sin(\Delta\phi_{j} + \beta) \bigg]. 
\end{eqnarray}
\end{subequations}

Hence, by exploiting the phase shift symmetry, the set of $2N$ equations~\eqref{eq:KMI_final_first_order_reduction} were reduced to a $2(N-1)$ dimensional system.
The elimination of two equations by a single symmetry arises from the second-order nature of the \acp{ode}. Since equations~\eqref{eq:KMI_with_all_parameters} are second-order \acp{ode} and exhibit a phase shift symmetry, introducing phase-difference coordinates removes one second-order \ac{ode}. After rewriting the system as a first-order system, this corresponds to the elimination of two first-order \acp{ode}.
\\
To summarize, the system of equations~\eqref{eq:KMI_with_all_parameters} is first transformed into a reference frame rotating with constant frequency $\omega$. By introducing the phase-difference coordinates, it is shifted into the co-rotating reference frame of oscillator 1, which itself is already situated in the rotating reference frame.

\subsubsection{Cluster Subspaces}
\label{subsec:cluster_subspaces}
The permutation symmetry of equations~\eqref{eq:KMI_with_all_parameters} implies the existence of invariant subspaces \cite{Ashwin.2025}. 
In the subspaces, at least two oscillators $i$ and $j$ are synchronized such that $\phi_i(t) = \phi_j(t), \ \dot{\phi}_i(t) = \dot{\phi}_j(t)$ for some $i \neq j$, $i,j \in \{1, \ldots , N\}$ and for all times $t \in \mathbb{R}$. 
In other words, once two oscillators are synchronous at one instance of time, they remain synchronous for all times.
We define an $n$-cluster subspace as the set of states with $n$ groups of synchronous oscillators, i.e., oscillators within a group have identical phases and phase velocities, whereas different groups may or may not be synchronized.\\
If a bifurcation involves only states that remain entirely within a given cluster subspace, it is referred to as a \textit{longitudinal bifurcation} in the following.
In contrast, if a bifurcation involves a branch that is not contained within the $n$-cluster subspace, it is referred to as a \textit{transversal bifurcation}.
\paragraph{2-Cluster subspace}
At first, we introduce the 2-cluster subspace, in which all oscillators are divided into two phase-synchronous groups. 
The fraction of oscillators in the first cluster is defined as $\rho_1\coloneqq\frac{N_1}{N}$, where $N_1$ is the number of oscillators in the first cluster. Without loss of generality, we assume that the first cluster is the larger one such that $\rho_1 \in [0.5, 1]$. With the phases $\phi_1,\phi_2$ and the phase velocities $\psi_1,\psi_2$ of the two clusters, the corresponding phase-difference coordinates are $\Delta \phi = \phi_1 - \phi_2$ and $\Delta \psi = \psi_1 - \psi_2$
and the 
equations for the cluster dynamics in the 2-cluster subspace are given by:
\begin{subequations}
\label{eq:KMI_2_cluster_phase_difference_equations}
    \begin{equation}
	\Delta \dot{\phi} = \Delta \psi
    \end{equation}
    \begin{eqnarray}
		\Delta \dot{\psi} = &&-\epsilon \cdot \Delta\psi
		- \sigma [\sin(\Delta \phi) \cos(\beta) \nonumber \\
        &&+ (2\rho_1 -1) \cdot \sin(\beta)  \cdot (1- \cos(\Delta \phi))] .
    \end{eqnarray}
\end{subequations}
Note that the equations do not depend on the number of oscillators $N$ and thus apply both to finite systems of arbitrary size and to the thermodynamic limit $N \rightarrow \infty$ for rational cluster distributions $\rho_1$.
\paragraph{3-Cluster subspace}
Analogously to the 2-cluster subspace, a 3-cluster subspace can be introduced in which now the oscillators are grouped into three phase-synchronous groups. With the sizes of the clusters $N_1,\ N_2,\ N_3 \ \mathrm{with}\ N_1+N_2+N_3=N$, the parameters $\rho_1 \coloneqq \frac{N_1}{N}$ and $\rho_2 \coloneqq \frac{N_2}{N_2 + N_3}$ can be defined. As before, $\rho_1$ describes the fraction of oscillators in the first cluster, and $\rho_2$ describes the fraction of oscillators in the second cluster, referred to the remaining oscillators in groups two and three. Without loss of generality, we assume that $N_1 \geq N_2 \geq N_3$ such that $\rho_1 \in [0.5, 1]$ and $\rho_2 \in [0.5, 1]$.
Introducing the phase-difference coordinates $\Delta\phi_{i}=\phi_1 - \phi_{i+1}$ and $\Delta\psi_{i}=\psi_1 - \psi_{i+1}$ with $i \in \{ 1,2 \}$, the governing equations read
\begin{subequations}
	\label{eq:KMI_3_cluster_phase_difference_equations}
\begin{equation}
	\Delta \dot{\phi}_{i} = \Delta \psi_{i}
\end{equation}
\begin{eqnarray}
		\Delta \dot{\psi}_{i} = &&-\epsilon \Delta\psi_{i} - \sigma \Big[\rho_1 \cdot (\sin(\beta) - \sin(-\Delta \phi_{i} + \beta) ) \nonumber\\
		&&+ (1-\rho_1) \rho_2 \cdot (\sin(\Delta \phi_{1}+\beta) - \sin(\Delta \phi_{1} - \Delta \phi_{i} + \beta) ) \nonumber\\
		&&+ (1-\rho_1) \cdot (1-\rho_2) \nonumber\\ 
        &&\cdot (\sin(\Delta \phi_{2}+\beta) - \sin(\Delta \phi_{2} -\Delta \phi_{i} + \beta) )\Big].
\end{eqnarray}
\end{subequations}

\subsection{Definition of Frequency Clusters}
To characterize collective behavior in the system, the concept of frequency clusters is introduced here. A single frequency cluster is defined as a group of oscillators that exhibit identical mean frequencies $\langle \dot{\phi} \rangle$. Frequency-cluster states, consisting of $n$ clusters ($n\geq2$) with distinct mean frequencies, are hereafter also referred to as $n$-cluster states. Frequency-cluster states are characterized by at least two diverging phase variables $\phi_i$ corresponding to one diverging phase difference $\Delta \phi_i$.
Two types of frequency-cluster states are distinguished. In an asynchronous frequency cluster, the oscillators share the same mean frequency but differ in their instantaneous phase (velocity) values. In contrast, a synchronous frequency cluster is characterized by equal phases of all oscillators within the cluster. Note that the cluster subspaces are suitable for studying the synchronous frequency-cluster states, but not for analyzing asynchronous frequency-cluster states.

\subsection{Numerical Methods}
\label{subsec:Numerical_methods}
\subsubsection{Solving \acp{ode}}
Numerical solutions of the \acp{ode} were obtained with \texttt{DifferentialEquations.jl} \cite{Rackauckas.2017}, a package implemented in Julia \cite{Bezanson.2017}. We use the solver \texttt{Tsit5} implementing a 5/4 Runge-Kutta method \cite{Tsitouras.2011}. During the transient period of 10000 time steps, we use adaptive time steps with both absolute and relative tolerances set to $10^{-12}$. Convergence to a limit cycle is assumed if corresponding maxima of the phase velocity (difference) across successive periods differ by at most $10^{-8}$. In the subsequent time span, in addition to adaptive time stepping with the specified tolerances, the solution is sampled at fixed time intervals of $dt=0.1$, and these data points are used for the analysis. 
We use random initial conditions and initialize the phase (difference) variables in the interval $[0,2\pi)$ and the frequency (difference) variables in $[-1,1]$.
\subsubsection{Continuing limit cycles}
For the continuation of a limit cycle in one parameter, the Julia package \texttt{BifurcationKit.jl} was utilized \cite{RomainVeltz.2020}. Here, the orthogonal collocation method is employed \cite{Dankowicz.2013,Doedel.1991}. As an initial guess for the continuation, a numerically obtained limit cycle is used. \\
For continuations of limit cycle solutions with diverging variables, as they occur in this work for some phase or phase differences, the usual periodicity conditions, which state that every variable must attain the same value at both the starting and ending points of the orbit, no longer hold. Since the phases and thereby also the phase differences are $2\pi$-periodic, it is allowed to shift the phase differences by a multiple of $2\pi$. 
The new periodicity conditions for the diverging phase variables $\Delta \phi_i$ are
\begin{equation}
	\label{eq:Periodicity_condition_modified}
	\Delta \phi_i (t=0)=\Delta \phi_i(t=T) + k_i \cdot 2\pi, 
\end{equation}
where $t$ is time, $T$ is the period of the orbit and $k_i \in \mathbb{Z}$ is a fixed multiple of $2\pi$, by which $\Delta \phi_i$ is shifted in one period. This periodicity condition was implemented in the file "PeriodicOrbitCollocation.jl" which can be found in the following path of the Julia installation: ".julia\textbackslash packages\textbackslash BifurcationKit\textbackslash yWo5k\textbackslash src\textbackslash periodicorbit\textbackslash Periodic\\OrbitCollocation.jl". This path applies to version 0.4.8 of \texttt{BifurcationKit.jl}. In this file, the periodicity condition is defined in the function \texttt{functional\_coll!} and can be modified accordingly.\\
Furthermore, the software AUTO-07p \cite{Doedel.1981} is used for continuations of limit cycles in two parameters in section~\ref{sec:Two_frequency_clusters}.
\\
Local bifurcations are identified in both software packages by calculating the Floquet multipliers. In contrast, branches of homoclinic bifurcations are computed with the HomCont routine implemented in AUTO-07p, which, in short, enforces an intersection of the unstable and stable manifolds of the corresponding fixed point.
\subsubsection{Numerical analysis of 2-cluster states}
\label{subsubsec:Numerical_methods_analysis_of_2_cluster_states}
For validating the bifurcation analysis of 2-cluster states with direct numerical simulations in section \ref{subsubsec:2_cluster_validation_with_numerics}, 
we solve the \acp{ode} for 100 equidistant values of $\beta$ within the interval $[0, 2]$. For every $\beta$ value, we 
employ 500 initial conditions close to synchronous 2-cluster states by choosing the frequencies $\dot{\phi}_1=-\rho_1 \sin(\beta)$ and $\dot{\phi}_2=(1-\rho_1) \sin(\beta)$ \cite{Munyayev.2024}, and the phase difference $\Delta\phi =\pi$ between the two clusters. We vary the cluster sizes by selecting all possible cluster ratios, $\rho_1 \in [0.5, 1]$, and evenly distributing the total number of runs across them. Furthermore, random noise of order $\mathcal{O}(10^{-8})$ is added to all variables.
From all runs at one $\beta$ value, we exclude those that did not converge to a state with two synchronous clusters exhibiting different frequencies.
The sizes of the clusters are determined by calculating the correlation of the phase velocities and grouping highly correlated oscillators. 
\subsubsection{Parameter ramp}
We refer to a parameter ramp as a method in which one starts from a numerically determined state and then changes the parameter adiabatically in small steps. For the new parameter value, the \acp{ode} are solved for a sufficiently long time such that all transients have decayed, using the end state as a new initial condition. 
In this work, step sizes of $\pm 10^{-3}$ were used. To test the state's stability, a perturbation can be added to the new initial condition in each step. Thus, parameter ramps provide an overview of the states and their stability as a parameter is changed. 

\section{Origin and Stability of Two Frequency Clusters in the Thermodynamic Limit \label{sec:2-freq}}
\label{sec:Two_frequency_clusters}

\begin{figure}[!h]
	\centering
	\includegraphics[width=0.45\textwidth]{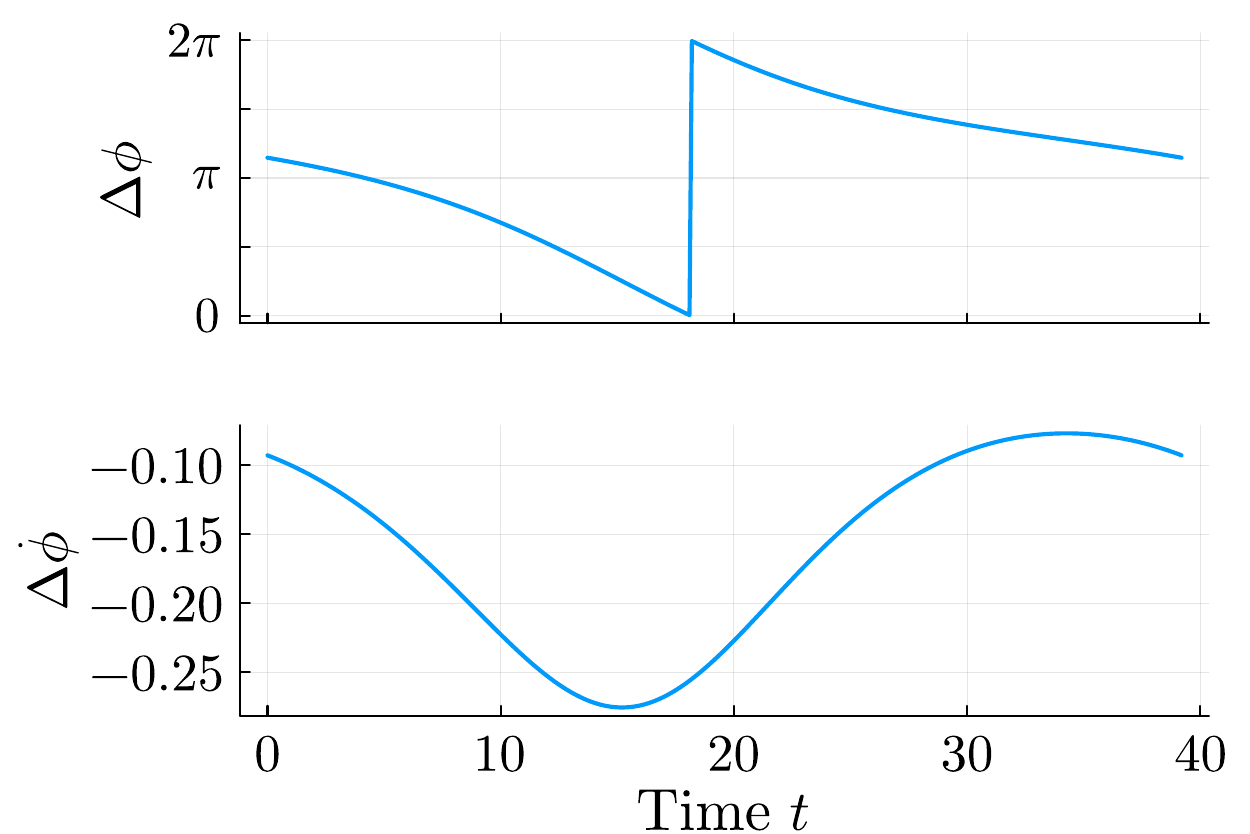}
	\caption{\justifying Two frequency clusters in the 2-cluster subspace described by equations~\eqref{eq:KMI_2_cluster_phase_difference_equations} containing one period. The time series of the phase difference $\Delta \phi$ and the frequency difference $\Delta \dot{\phi}$ are plotted. The parameter values $\rho_1=0.6$ and $\beta=0.4\pi$ were used.}
	\label{fig:2_cluster_state_temporal_evolution}
\end{figure}
\begin{figure}[!h]
    \centering
    \includegraphics[width=0.45\textwidth]{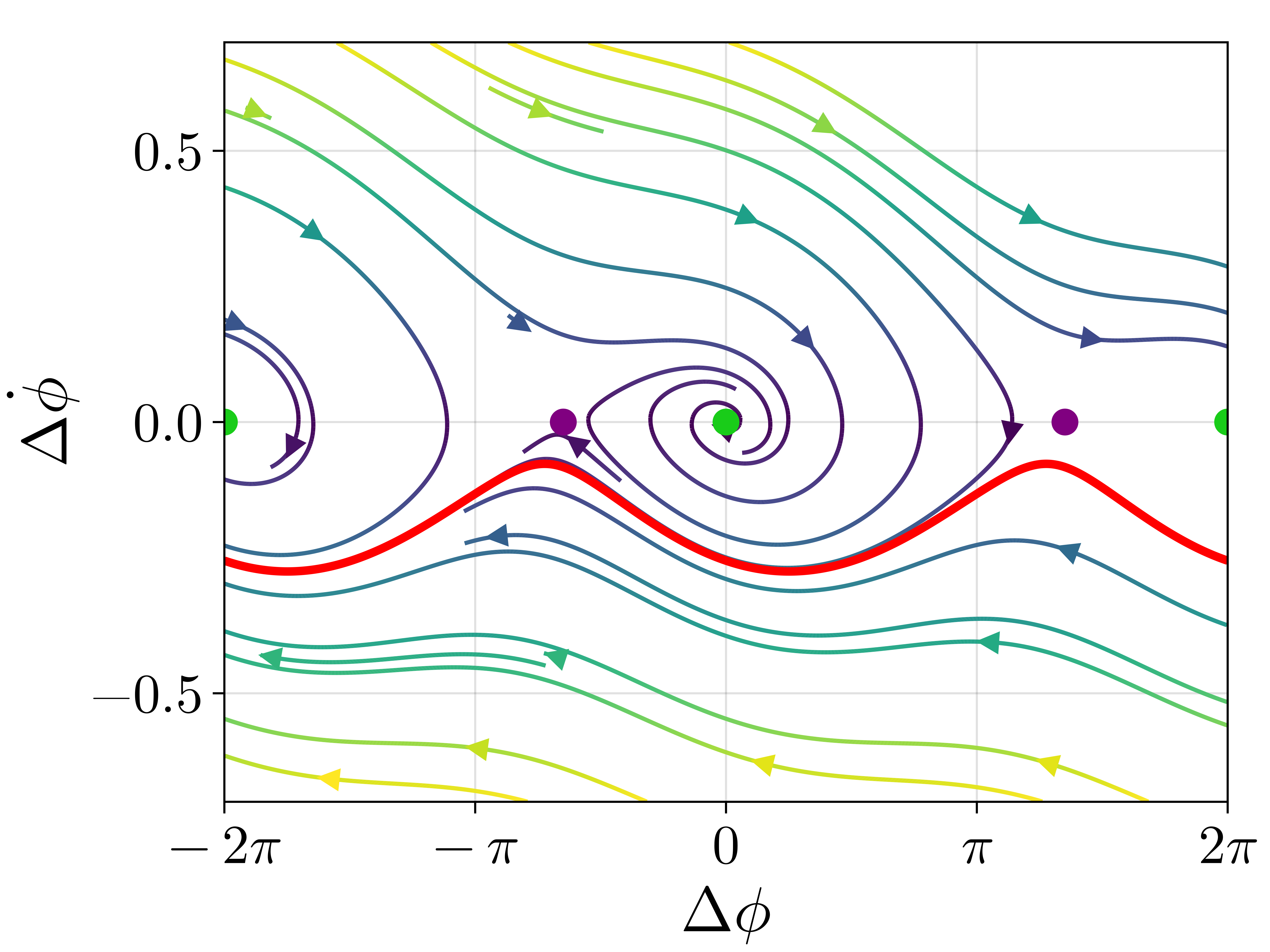}
    \caption{\justifying Phase space of the 2-cluster subspace from equations~\eqref{eq:KMI_2_cluster_phase_difference_equations} for $\rho_1=0.6$ and $\beta=0.4\pi$. 
    The phase-synchronous and two-cluster fixed points are represented by the green and purple dots, respectively, while the red line represents the limit cycle of two frequency clusters.}
    \label{fig:phase_space_2_cluster_subspace}
\end{figure}

In this section, we report on the origin and the stability of two synchronous frequency clusters, which have been found and investigated in finite-size systems of the \ac{kmi} \cite{Maistrenko.2017, Brezetsky.2021}. We go to the thermodynamic limit and extend previous work on stability regions \cite{Munyayev.2024} by investigating the bifurcations that destabilize two frequency clusters. To this end, we conduct a numerical bifurcation analysis using the cluster subspaces from section~\ref{subsec:cluster_subspaces}. First, we investigate the longitudinal stability of 2-cluster states in section \ref{subsec:two_cluster_stability_1_param}. Second, we investigate their transversal stability in section \ref{subsec:2_cluster_transversal_stability} before summarizing their stability in figure \ref{fig:2_cluster_heatmap_100_osc_adapted_ics}.\\
In figure~\ref{fig:2_cluster_state_temporal_evolution}, a stable two-frequency-cluster state is shown that was obtained employing the 2-cluster subspace of the \ac{kmi}  (equations~\eqref{eq:KMI_2_cluster_phase_difference_equations} with $\rho_1=0.6$). Here, the temporal evolution of the phase difference $\Delta\phi$ and the frequency difference $\Delta\dot{\phi}=\Delta\psi$ between the two clusters is plotted over one period of the cycle. 
This state is characterized by a non-vanishing average of the frequency difference $\langle \Delta\dot{\phi}\rangle$ and therefore a diverging phase difference $\Delta \phi$ which we restrict to $\Delta\phi \in [0,2\pi)$. This state can also be visualized in phase space, which is done in figure~\ref{fig:phase_space_2_cluster_subspace}, where the red line represents two frequency clusters. From both figures~\ref{fig:2_cluster_state_temporal_evolution} and \ref{fig:phase_space_2_cluster_subspace}, 
it is evident that the mean frequency difference between the clusters is nonzero, which implies distinct cluster frequencies and thus a rotational motion, i.e., a rotation frequency, of the phase difference. This should be distinguished from a libration frequency, which refers to the frequency of oscillatory motion confined to a bounded region in phase space that, for example, occurs in the non-rotational motion of an undamped pendulum.

\subsection{Fixed Points}
\label{subsec:two_clusters_fixed_points}
In this section, we study the fixed points of the 2-cluster subspace in phase-difference coordinates, as defined by equations~\eqref{eq:KMI_2_cluster_phase_difference_equations}. We only report fixed points that will be involved in the bifurcations of frequency clusters. \\
The synchronous state, given by $\Delta \phi=0$, $\Delta \psi=0 \ $\cite{Munyayev.2024}, is always a fixed point, regardless of the parameter values.  It is stable for $\beta \in [0,\frac{\pi}{2})$ and unstable for $\beta \in (\frac{\pi}{2}, \pi]$. In the non-rotating laboratory reference frame of the \ac{kmi} (cf. equations~\eqref{eq:KMI_with_all_parameters}), it corresponds to a phase-synchronous limit cycle, which rotates with $\dot{\phi} = \omega - \frac{\sigma}{\epsilon}\sin(\beta)$ .\\
Furthermore, there is a two-cluster fixed point describing two groups with the phase difference 
\begin{equation}
\label{eq:fixed_point_two_clusters_phase_difference}
    \Delta \phi = 2 \arctan \left( \frac{\cot(\beta)}{1-2\rho_1} \right)
\end{equation}
and $\Delta \psi=0 \ $\cite{Munyayev.2024} for $\rho_1 \neq 0.5$. 
The state is unstable for $\beta \in [0,\frac{\pi}{2})$ and stable for $\beta \in (\frac{\pi}{2}, \pi]$. At $\beta=\frac{\pi}{2}$, both fixed points undergo a transcritical bifurcation, exchanging their stabilities.

\subsection{Longitudinal Stability and Origin of Two Frequency Clusters}
\label{subsec:two_cluster_stability_1_param}
\begin{figure}[!h]
	\centering
	\includegraphics[width=0.45\textwidth]{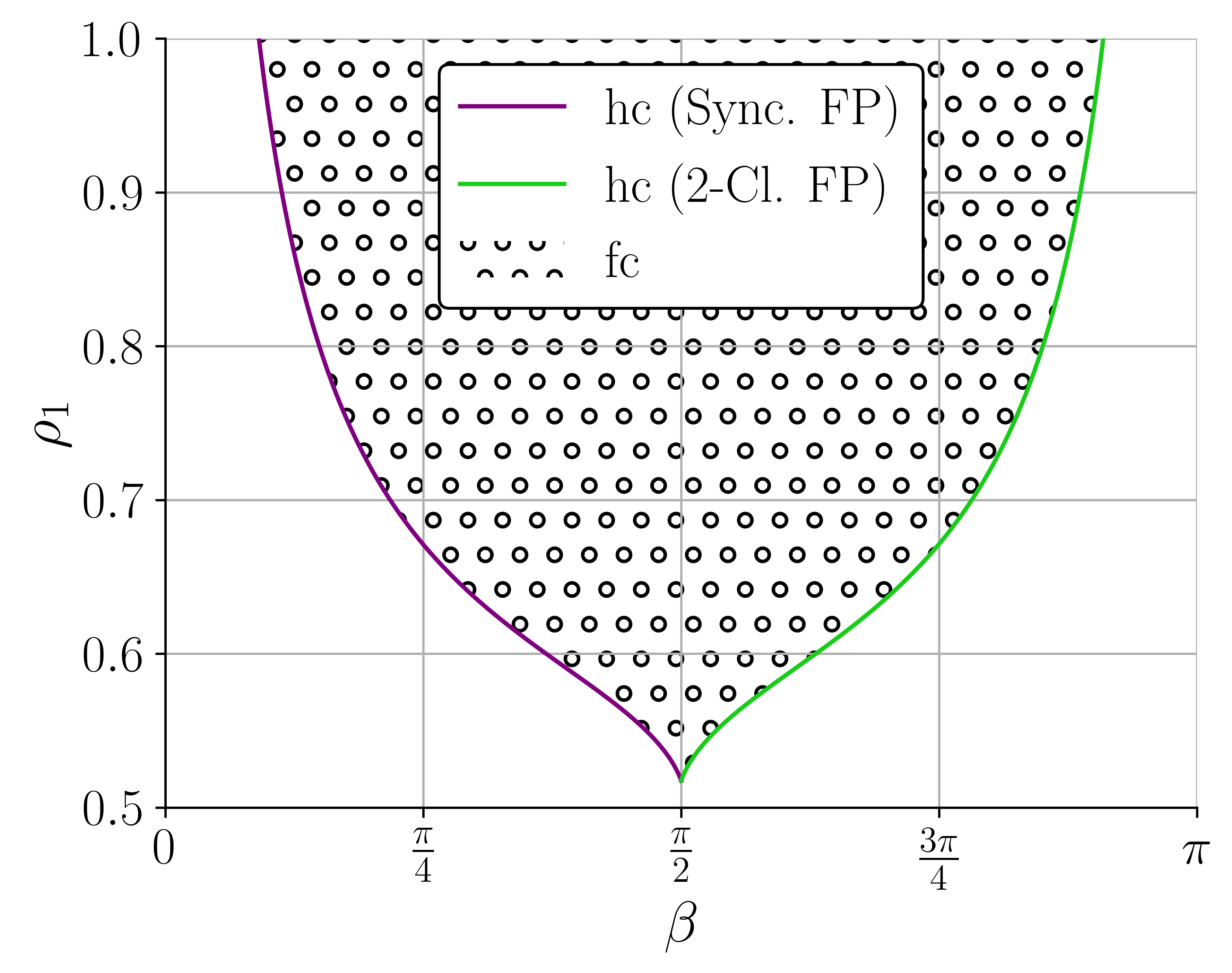}
	\caption[Homoclinic bifurcations in the $\beta-\rho_1$ plane of the 2-cluster subspace]{\justifying Longitudinal bifurcations of two synchronous frequency clusters (fc) in the $\beta-\rho_1$ parameter plane of the 2-cluster subspace of the \ac{kmi}, described by equations~\eqref{eq:KMI_2_cluster_phase_difference_equations}. The area with the circular pattern represents the region where 2-cluster states are longitudinally stable. The colored lines highlight the homoclinic bifurcations (hc) involving the corresponding fixed point (FP) that is shown in the legend.
    }
	\label{fig:Homocline_positions_in_2_cluster_subspace}
\end{figure}

To begin the analysis, we first recapitulate the findings on the longitudinal stability of two frequency clusters in the thermodynamic limit as reported by Munyayev et al. \cite{Munyayev.2024}. We present these results in a manner suitable for the subsequent analysis of their transversal stability.\\
We determine the longitudinal stability by restricting the system to the 2-cluster subspace according to equations~\eqref{eq:KMI_2_cluster_phase_difference_equations}. This means that the oscillators are grouped into two internally phase- and frequency-synchronous clusters with a fixed cluster size ratio $\rho_1$. 
Thus, only those bifurcations that do not split the clusters can be captured, which are by definition the longitudinal bifurcations.\\
We start by studying the bifurcations of the synchronous two frequency clusters, an example of which is shown in figure~\ref{fig:2_cluster_state_temporal_evolution}. 
The two frequency clusters exist in a $\beta$ interval that is symmetric around $\beta =\pi / 2$, and they emerge and are destroyed in a homoclinic bifurcation. Figure~\ref{fig:Homocline_positions_in_2_cluster_subspace} shows the locations of the homoclinic bifurcations in the $\beta-\rho_1$ parameter plane. 
Within the region marked with open circles, which is bounded by the homoclinic bifurcations, longitudinally stable synchronous 2-cluster states exist. For $\beta < \frac{\pi}{2}$, the homoclinic orbit originates from the two-cluster fixed point, while for $\beta > \frac{\pi}{2}$, the homoclinic orbit emerges from the synchronous fixed point. At $\beta=\frac{\pi}{2}$, both branches merge. At this value of $\beta$, the synchronous fixed point and the two-cluster fixed point undergo a transcritical bifurcation, exchanging their stability.\\
As can be seen in figure~\ref{fig:Homocline_positions_in_2_cluster_subspace}, for large values of $\rho_1$, which corresponds to a strongly asymmetric distribution of oscillators into two groups, frequency clusters exist in a comparatively wide $\beta$-interval. Decreasing $\rho_1$ towards 0.5, i.e., approaching a balanced state with two groups of equal size, the $\beta$-range where frequency clusters exist shrinks. In our continuation, where we employed the values $\sigma=\epsilon=0.05$,
the homoclinic branches merge at $\rho_1\approx0.518$, which is the minimal value of $\rho_1$ for which frequency clusters exist. This is in agreement with results obtained for the \ac{kmi} which showed that states with two frequency clusters of equal sizes do not exist \cite{Munyayev.2024}.
\\
Figure~\ref{fig:Homocline_positions_in_2_cluster_subspace} also reveals the coexistence of multiple 2-cluster states with different group sizes for a given value of $\beta$. For instance, in a finite system of 10 oscillators, the 2-cluster state with $\rho_1=0.6$, which corresponds to cluster sizes 6 and 4, coexists for $\beta=\pi/2$ with the more asymmetric 2-cluster states with $\rho_1=0.7$, $\rho_1=0.8$, and $\rho_1=0.9$. All 2-cluster states also coexist with the phase-synchronous fixed point, describing a 1-cluster state with one common frequency, which is stable for $\beta \in [0,\frac{\pi}{2})$.
In summary, we have confirmed that two frequency clusters in the 2-cluster subspaces are created and destroyed by homoclinic bifurcations~\cite{Belykh.2016}, from which we conclude that the origin of two frequency clusters in the \ac{kmi} are homoclinic bifurcations.

\subsection{Transversal Stability of Two Frequency Clusters}
\label{subsec:2_cluster_transversal_stability}
As we already stated in the previous section, when restricting the dynamics to the 2-cluster subspace described by equations~\eqref{eq:KMI_2_cluster_phase_difference_equations}, only the longitudinal stability can be investigated. To study the transversal stability of two frequency clusters, we use equations~\eqref{eq:KMI_3_cluster_phase_difference_equations}, which describe the dynamics in the 3-cluster subspace. By setting $\rho_2=1$, we assign the third cluster zero weight. Consequently, this "test cluster" does not influence the dynamics of the two primary clusters, but its own evolution remains governed by them. When we initialize the test cluster synchronized with one of the two clusters, it can separate from the corresponding cluster, allowing us to assess the stability of that cluster.
At first, we investigate the stability of the smaller cluster with a numerical continuation.
Subsequently, we apply the same approach to the larger cluster to determine the bifurcations it experiences.
\subsubsection{Stability of the small cluster}
\begin{figure}[!h]
	\centering
	\includegraphics[width=0.45\textwidth]{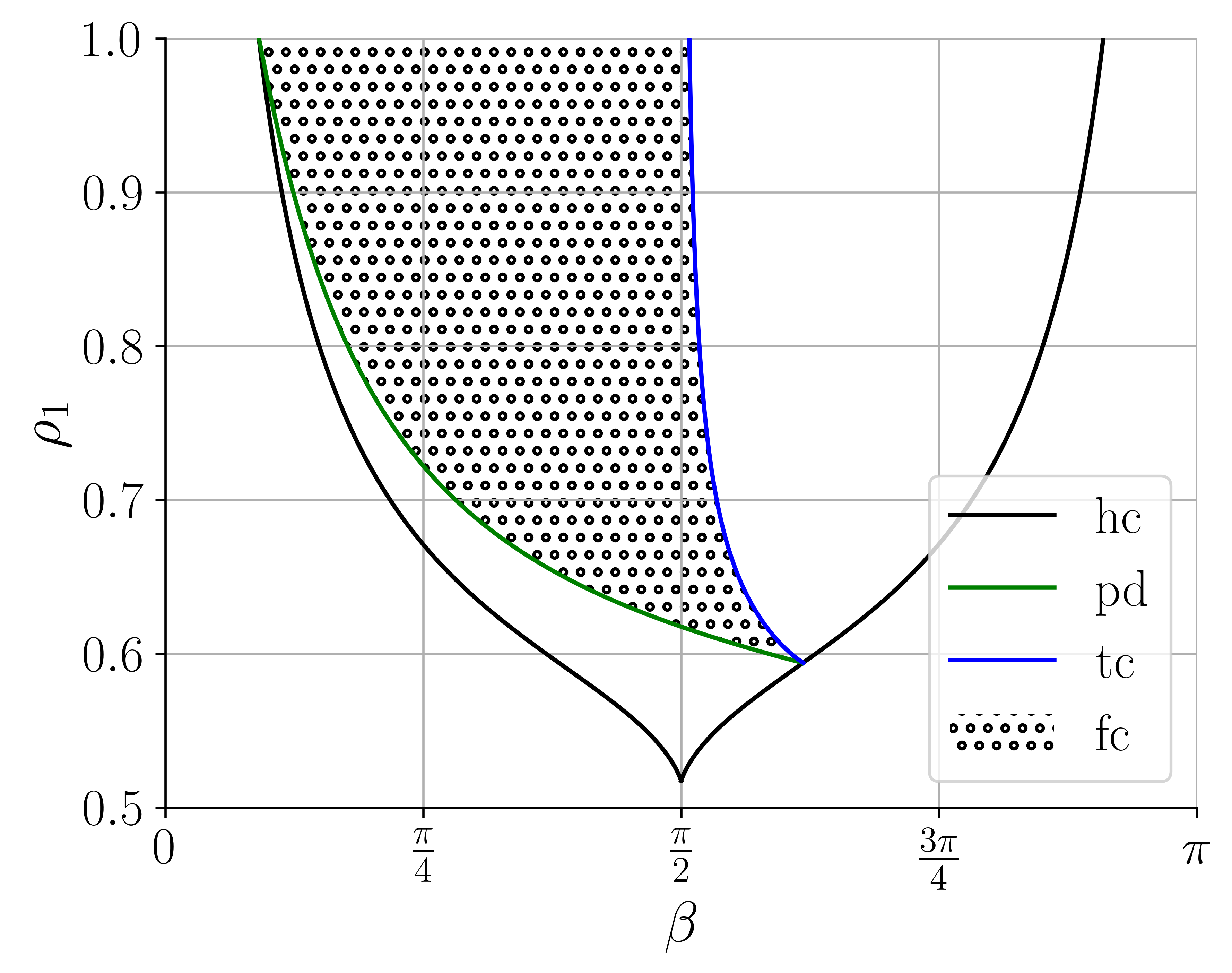}
	\caption{\justifying Transversal bifurcations of the small cluster in the $\beta-\rho_1$ parameter plane of the 2-cluster subspace. The area with the circular pattern indicates the region of synchronous frequency clusters (fc) with a stable small cluster. The bifurcation branches are illustrated by the colored lines. The abbreviation hc denotes a homoclinic bifurcation, pd a period-doubling bifurcation, and tc a transcritical bifurcation.
    }
	\label{fig:2_clusters_transversal_stability_small_cluster}
\end{figure}

The initial condition $\Delta \phi_2=\Delta \phi_1$ and $\Delta \dot{\phi}_2=\Delta \dot{\phi}_1$ places the test oscillator on the smaller cluster, whose stability is investigated in this section.
The result of a numerical bifurcation analysis of two frequency clusters 
in the $\rho_1$ and $\beta$ parameter plane is shown in figure~\ref{fig:2_clusters_transversal_stability_small_cluster}. Here, the region where the frequency clusters are stable is highlighted with a circular pattern. Towards smaller and larger $\beta$ values the frequency clusters are destabilized by period doubling and transcritical bifurcations, respectively. These two branches coalesce and interact with the homoclinic bifurcation in a codimension-2 point at $\beta\approx 1.9427$, $\rho_1\approx0.5941$.
\subsubsection{Stability of the large cluster}
\begin{figure}[!h]
	\centering
	\includegraphics[width=0.45\textwidth]{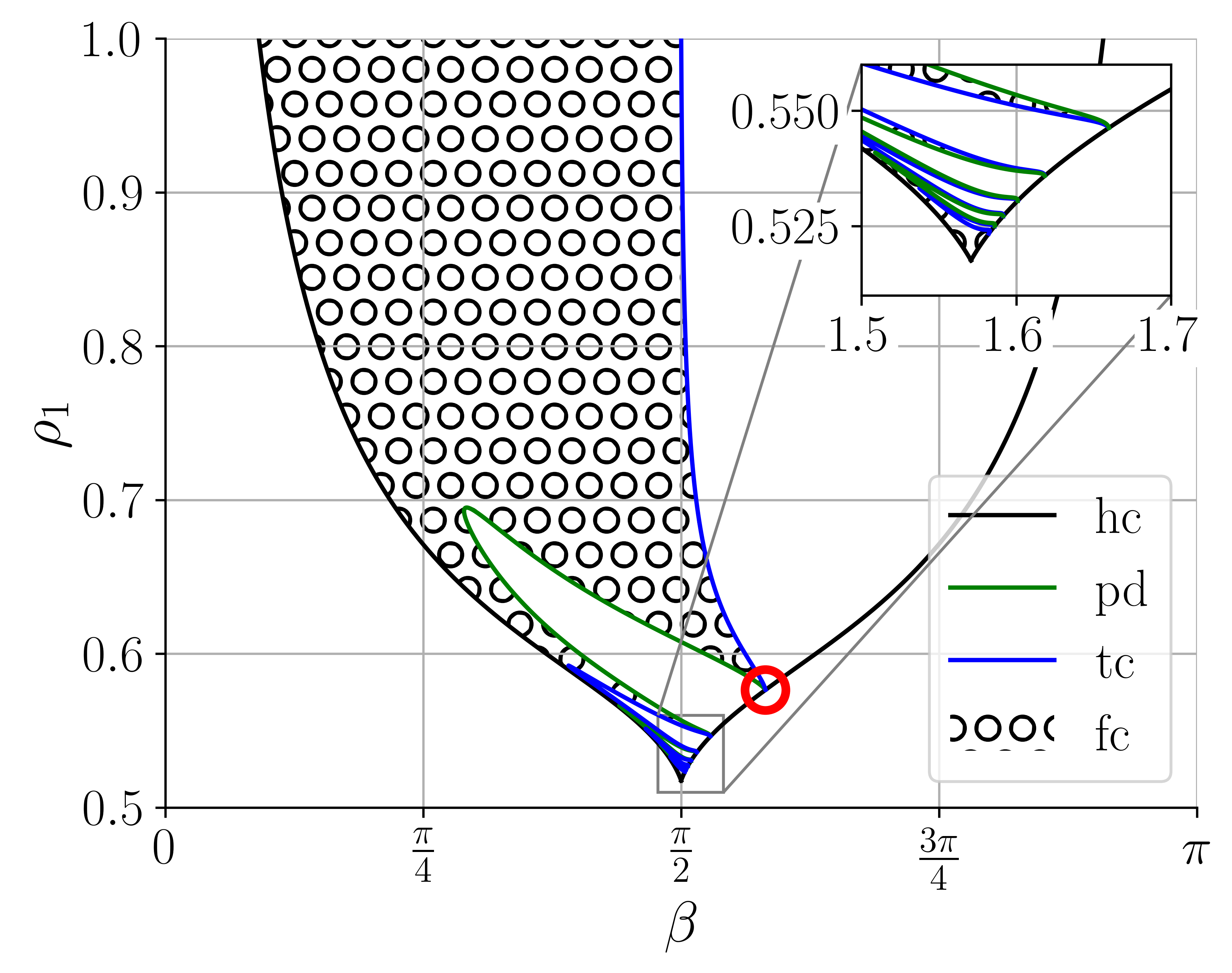}
	\caption{\justifying Transversal bifurcations of the large cluster in the $\beta-\rho_1$ parameter plane of the 2-cluster subspace. The area with the circular pattern indicates the region of synchronous frequency clusters (fc) with a stable large cluster. The bifurcation branches are illustrated by the colored lines and the red circle highlights the first codimension-2 point in a sequence of such points. The abbreviation hc denotes a homoclinic bifurcation, pd a period-doubling bifurcation, and tc a transcritical bifurcation.
    }
	\label{fig:2_clusters_transversal_stability_large_cluster}
\end{figure}

Next, we study the stability of the large cluster. To this end, we initialize the test cluster synchronized with the large cluster by setting $\Delta \phi_2 = 0$ and $\Delta \dot{\phi}_2 = 0$. As before, we continue the 2-cluster state in two parameters. The resulting bifurcation diagram is shown in figure~\ref{fig:2_clusters_transversal_stability_large_cluster}. The large cluster is transversely stable in the area filled with open circles. At the lower $\beta$ border of this region, the stability range extends up to the homoclinic bifurcation.
Coming from the most asymmetric clusters, the stability boundary at large $\beta$ values is bound by a transcritical bifurcation that ends in a codimension-2 point, highlighted with a red circle, where it interacts with the homoclinic and a transverse period-doubling bifurcation, just as we observed it above for the smaller cluster (see Fig.~\ref{fig:2_clusters_transversal_stability_small_cluster}). However, the transverse period doubling bifurcation does not form the low-$\beta$ boundary of the stability region. Instead, it extends only to $\rho_1\approx0.7$ where it bends towards smaller $\rho_1$ values again, to interact a second time in a codimension-2 bifurcation with a transcritical and the homoclinic bifurcation. From there a similar, yet much smaller loop of transcritical bifurcations links to a further codimension-2 point on the homoclinic bifurcation at still lower values of $\rho_1$ (see the inset of Fig.~\ref{fig:2_clusters_transversal_stability_large_cluster}). These subsequent tongues of period doubling and transcritical bifurcations seem to occur a few times in ever smaller parameter regions
, approaching $\beta=\frac{\pi}{2}$ asymptotically.
Remarkably, this bifurcation pattern was also observed recently in a network of mean-field coupled Stuart-Landau oscillators \cite{Thome.2025}. 
In this work, the stability of states with two amplitude clusters was investigated. In both the Stuart-Landau model and the \ac{kmi}, the limit cycle (in phase difference coordinates) describing two phase-synchronous clusters loses its stability through the mentioned bifurcation pattern. As a result, new limit cycles emerge that represent states in which an individual cluster is no longer phase-synchronous.
Furthermore, in a bifurcation study of three identical and globally coupled Kuramoto oscillators with inertia, a codimension-2 bifurcation was reported, in which, analogously to our study, a transcritical, a period-doubling and a homoclinic bifurcation merge\cite{Ashwin.2025}. This organizing center seems to correspond to our first codimension-2 point at $\rho_1 \approx 0.5764$, $\beta \approx1.8275$, which is highlighted by a red circle (cf. figure \ref{fig:2_clusters_transversal_stability_large_cluster}).

\subsubsection{States emerging in transversal bifurcations}
\begin{figure}[!h]
	\centering
	\includegraphics[width=0.45\textwidth]{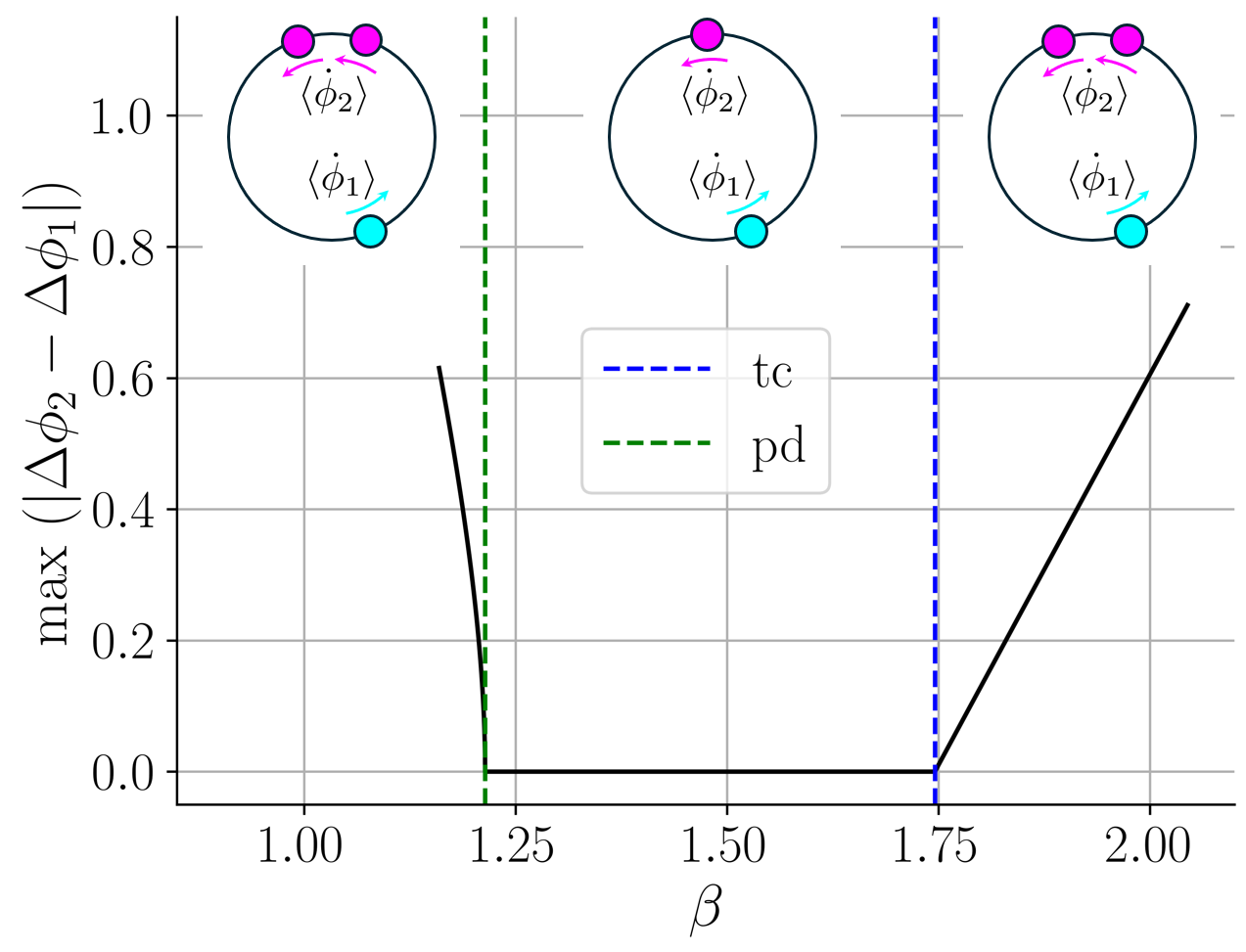}
	\caption{\justifying Phase desynchronization of the smaller frequency cluster for fixed cluster size ratio $\rho_1=0.65$, cf. figure~\ref{fig:2_clusters_transversal_stability_small_cluster}. The maximum of the absolute phase difference between the small cluster and the test cluster $\text{max }(|\phi_2 - \phi_3|)=\text{max }(|\Delta\phi_2 - \Delta\phi_1|)$ over one period is displayed. This maximum is shown for the states emerging from the period-doubling bifurcation (pd, left), from the transcritical bifurcation (tc, right), and for the state with phase-synchronous clusters (middle). The upper sketches depict the phase variables and illustrate the loss of phase synchrony of the small (pink) cluster.}
	\label{fig:2_cluster_asynchronous_states}
\end{figure}

In the previous part, we have shown that each of the two phase-synchronous frequency clusters can be destabilized by two distinct transversal bifurcations: a period-doubling bifurcation or a transcritical bifurcation. Numerical explorations of finite systems suggest that states emerging in transcritical bifurcations are presumably transversally unstable for $N>3$, while period-doubled cycles are stable. 
We characterize these states emerging in the bifurcations of the small cluster in figure~\ref{fig:2_cluster_asynchronous_states}, which are qualitatively equal to the states emerging from the large cluster's bifurcations. Here, we continue the states in $\beta$ and plot the maximal phase difference between the small cluster and the test cluster. The states emerging from both bifurcations, indicated by the dashed lines, exhibit a non-vanishing phase difference between the small cluster and the test cluster. Therefore, both bifurcations break the phase-synchrony of the corresponding cluster. However, this does not influence the mean frequency which is still equal within the phase-asynchronous cluster. According to our definition, this still constitutes a frequency-cluster state. Thus, asynchronous frequency-cluster states bifurcate from synchronous frequency-cluster states.

\subsubsection{Validation with numerical data}
\label{subsubsec:2_cluster_validation_with_numerics}
\begin{figure}[!h]
	\centering
	\includegraphics[width=0.45\textwidth]{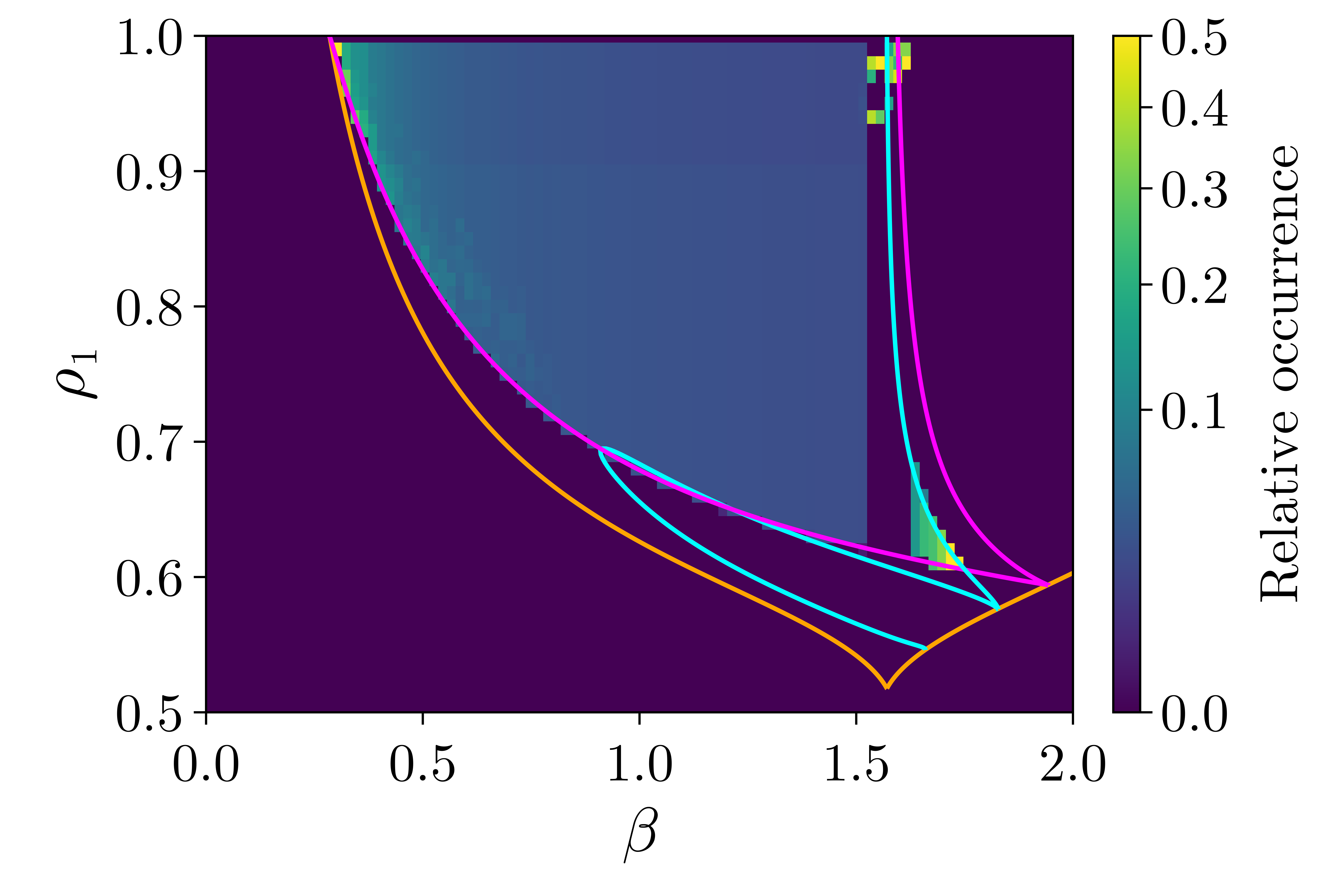}
	\caption{\justifying Numerical simulation of a network of 100 oscillators showing the relative occurrence of phase-synchronous 2-cluster states for a given value of $\beta$. The initial conditions are chosen close to the 2-cluster states. 
    Pink and turquoise lines indicate transversal bifurcations destabilizing the small and large cluster, respectively, while orange lines denote homoclinic bifurcations, cf. figures~\ref{fig:2_clusters_transversal_stability_small_cluster} and \ref{fig:2_clusters_transversal_stability_large_cluster}. Colors are scaled using a square-root mapping, saturating values larger than 0.5.}
	\label{fig:2_cluster_heatmap_100_osc_adapted_ics}
\end{figure}

To validate our approach of restricting the system to the cluster subspaces and analyzing the cluster stabilities with the test cluster, we numerically solve the \acp{ode} from equations~\eqref{eq:KMI_final_first_order_reduction} for a system of 100 oscillators.
For details regarding the numerics, we refer the reader to section \ref{subsubsec:Numerical_methods_analysis_of_2_cluster_states}.
The result of the computation is shown in figure~\ref{fig:2_cluster_heatmap_100_osc_adapted_ics}. Here, the multistability of 2-cluster states with different cluster sizes is again clearly visible. In almost the complete region, where stable 2-cluster states are predicted, bounded by the colored bifurcation branches, 2-cluster states are observed. Only in the region close to the transcritical bifurcations near $\beta=\frac{\pi}{2}$, no 2-cluster states are found. Most likely, the basin of attraction becomes very small. Furthermore, some 2-cluster states are detected outside of the predicted region for large $\rho_1$ at $\beta \approx 1.7$. These states are likely to be long transients. Overall, the numerical result agrees very well with the bifurcation analysis within the cluster subspace, thereby justifying the approach of reducing the system to its subspaces and employing a test cluster.

\section{Three Frequency Clusters in a System with 7 Oscillators}
\label{sec:3_frequency_clusters}
\begin{figure}[!h]
	\centering
	\includegraphics[width=0.45\textwidth]{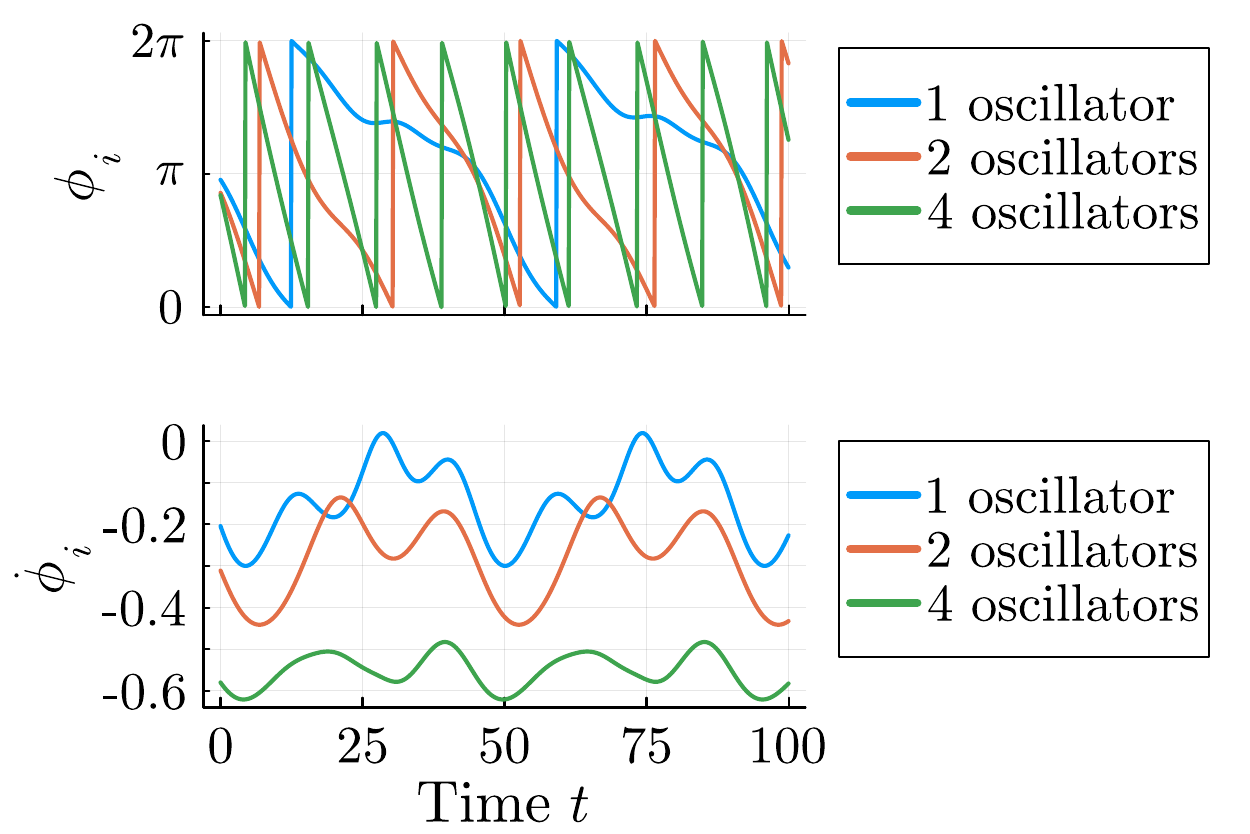}
	\caption{\justifying Time series of a state with three frequency clusters in a network of 7 oscillators described by equations~\eqref{eq:KMI_with_all_parameters} with $\beta=1.4$. The temporal evolution of the oscillators' phases $\phi_i$ and frequencies $\dot{\phi}$ is displayed. The legend shows the size of the clusters.} 
	\label{fig:7_osc_3_clusters_time_series}
\end{figure}

In the previous section, we showed that two frequency clusters originate from homoclinic bifurcations. Now, we extend our study of frequency clusters in the \ac{kmi} to the emergence of three frequency clusters. Numerical studies reveal that three frequency clusters emerge in a system with seven oscillators. Such a state is depicted in figure~\ref{fig:7_osc_3_clusters_time_series}. Here, the evolution of the oscillators' phases $\phi_i$ and frequencies $\dot{\phi}_i$ is shown. The green curve represents four synchronous oscillators, while the orange curve represents two, and the blue curve a single oscillator. This 4-2-1 state therefore consists of three clusters, each with a different mean frequency. 

In the following, we investigate this 3-cluster state by analyzing its longitudinal and transversal stability using numerical continuation. 

\subsection{Longitudinal Stability and Origin of Three Frequency Clusters}
In this section, we examine the longitudinal stability of the three frequency clusters from figure~\ref{fig:7_osc_3_clusters_time_series} in the 3-cluster subspace using equations~\eqref{eq:KMI_3_cluster_phase_difference_equations}. Setting $\rho_1=\frac{4}{7}$ and $\rho_2=\frac{2}{3}$ yields the desired ratio between the cluster sizes of 4:2:1. 
%
Recall that by employing the 3-cluster subspace, we fix the oscillators into three groups and do not allow them to break up, which is the same principle as in section~\ref{sec:Two_frequency_clusters} for two clusters. Furthermore, the equations for the 3-cluster subspace also apply in the thermodynamic limit, meaning that the analysis of the longitudinal stability also holds for arbitrary system sizes, as long as the number of oscillators is $N=k \cdot 7$ oscillators, where $k \in \mathbb{N_+}$ such that a cluster ratio of 4:2:1 can be fulfilled.
At first, we show the 3-cluster state from figure~\ref{fig:7_osc_3_clusters_time_series} transformed into phase-difference coordinates within the 3-cluster subspace. Figure~\ref{fig:7_osc_subspace_4_2_1_state_start_cycle} displays the temporal evolution of the phase differences $\Delta \phi_i$ and the frequency differences $\Delta \dot{\phi}_i$. Here, an oscillator from the group with 4 oscillators is used as a reference oscillator, which is the consequence of choosing $\rho_1=\frac{4}{7}$. In the time series, a single period of the limit cycle is illustrated. 
All variables attain the same values in the beginning and the end of the cycle.
The phase differences are restricted to $\Delta \phi_i \in [0,2\pi)$ by performing mod $2\pi$ operations. Therefore, the phase differences actually undergo shifts of $\Delta \phi_1(t=T) - \Delta \phi_1 (t=0) = -2 \cdot 2\pi$ and $\Delta \phi_2(t=T) - \Delta \phi_2 (t=0) = -3 \cdot 2\pi$ during one period $T$, which is visible from the jumps in the phase differences. Diverging phase differences occur if and only if the system exhibits frequency clusters. The ratio of the shifts in phase differences within one period, which is given by $\frac{\Delta \phi_1(t=T) - \Delta \phi_1 (t=0)}{\Delta \phi_2(t=T) - \Delta \phi_2 (t=0)} = \frac{2}{3}$, determines the ratio of the average frequency differences, which attain the same value $\frac{\langle\Delta\dot{\phi}_1\rangle}{\langle\Delta\dot{\phi}_2\rangle} = \frac{2}{3}$.\\
One can generalize this locking of the mean frequency differences as 
\begin{equation}
\label{eq:locking_of_phase_differences}
    \frac{\langle\Delta\dot{\phi}_1\rangle}{\langle\Delta\dot{\phi}_2\rangle} = \frac{\langle \dot{\phi}_1 \rangle - \langle \dot{\phi}_2 \rangle}{\langle \dot{\phi}_1 \rangle - \langle\dot{\phi}_3\rangle} = \frac{k}{l} \quad \text{for } k,l \in \mathbb{Z},\, k\neq l
\end{equation}
to express an arbitrary rational ratio with the integers $k$ and $l$. Rewriting this equation yields
\begin{equation}
    \label{eq:triplet_locking_condition}
    (l-k) \langle\dot{\phi}_1\rangle - l \langle\dot{\phi}_2\rangle + k \langle\dot{\phi}_3\rangle = 0,
\end{equation}
which is a condition that characterizes a state exhibiting triplet synchrony or triplet locking, as described by Kralemann et al. \cite{Kralemann.2013}. We see that a usual locking of two phase differences, given by equation~\eqref{eq:locking_of_phase_differences} translates into a triplet-locking of the phase variables as in equation~\eqref{eq:triplet_locking_condition}.\par
\begin{figure}[!h]
	\centering
	\includegraphics[width=0.45\textwidth]{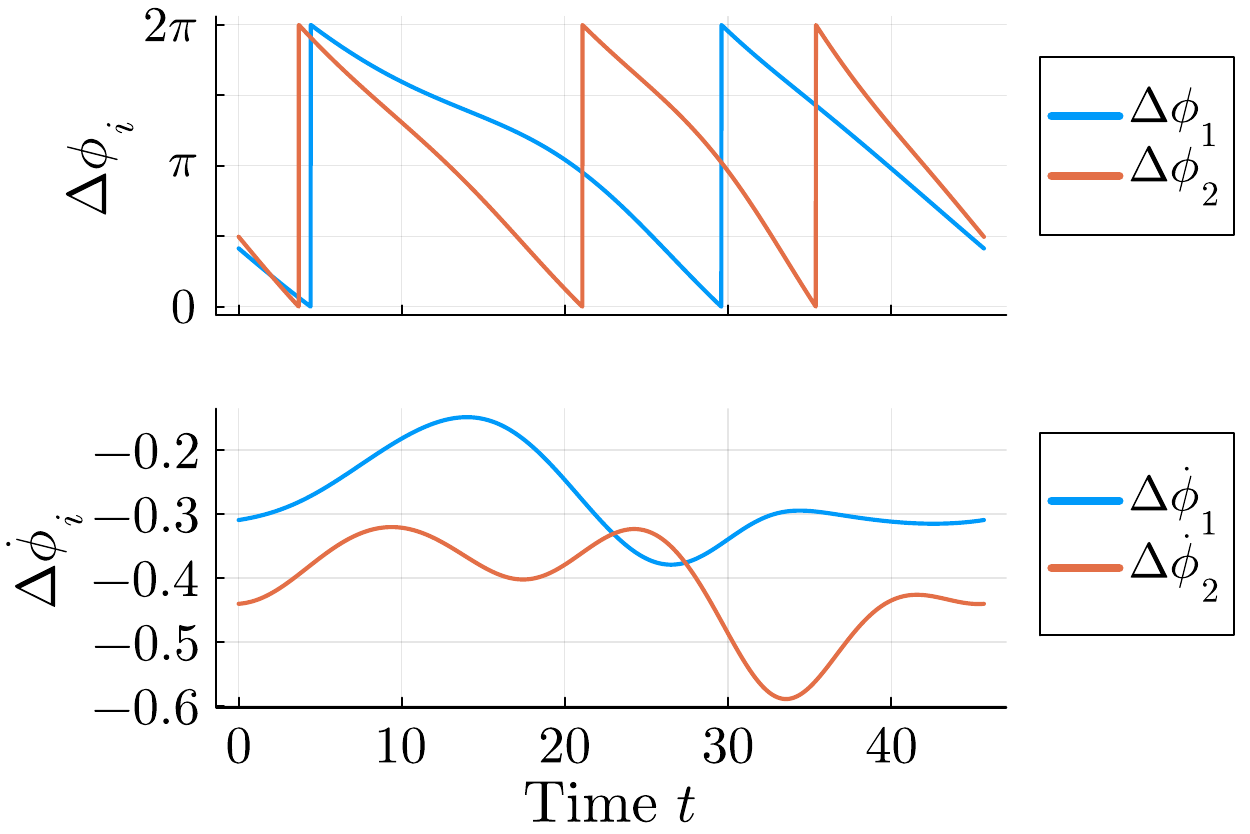}
	\caption[4-2-1 state in the 3-cluster subspace]{\justifying 4-2-1 frequency-cluster state in the 3-cluster subspace described by equations~\eqref{eq:KMI_3_cluster_phase_difference_equations} with $\beta=1.4$, $\rho_1=\frac{4}{7}$ and $\rho_2=\frac{2}{3}$, containing one period. In the upper plot, the time series of the phase differences $\Delta \phi_i$ is shown, and in the lower plot, the frequency differences $\Delta \dot{\phi}_i$ are plotted.}
	\label{fig:7_osc_subspace_4_2_1_state_start_cycle}
\end{figure}
Now, we perform a continuation of the frequency-cluster state from figure~\ref{fig:7_osc_subspace_4_2_1_state_start_cycle} in the parameter $\beta$. The result is displayed in figure~\ref{fig:7_osc_subspace_4_2_1_state_continuation}, where the maximum of $\Delta\dot{\phi}_2$ within one cycle is plotted on the ordinate.
Solid lines represent stable cycles, whereas dashed lines represent unstable cycles. The coloured dots indicate locations at which bifurcations occur. The periodic orbit is stable for $\beta \in [1.3296, 1.6877]$. At $\beta \approx 1.3296$, the state is destabilized by a period-doubling bifurcation, and at $\beta \approx 1.6877$, the stable branch undergoes a saddle-node bifurcation of limit cycles.
Additional noteworthy bifurcations occur at the endpoints of the continuation. At both points $\beta \approx 0.9300$ and $\beta \approx 1.1365$, homoclinic bifurcations arise.
\begin{figure}[!h]
	\centering
	\includegraphics[width=0.45\textwidth]{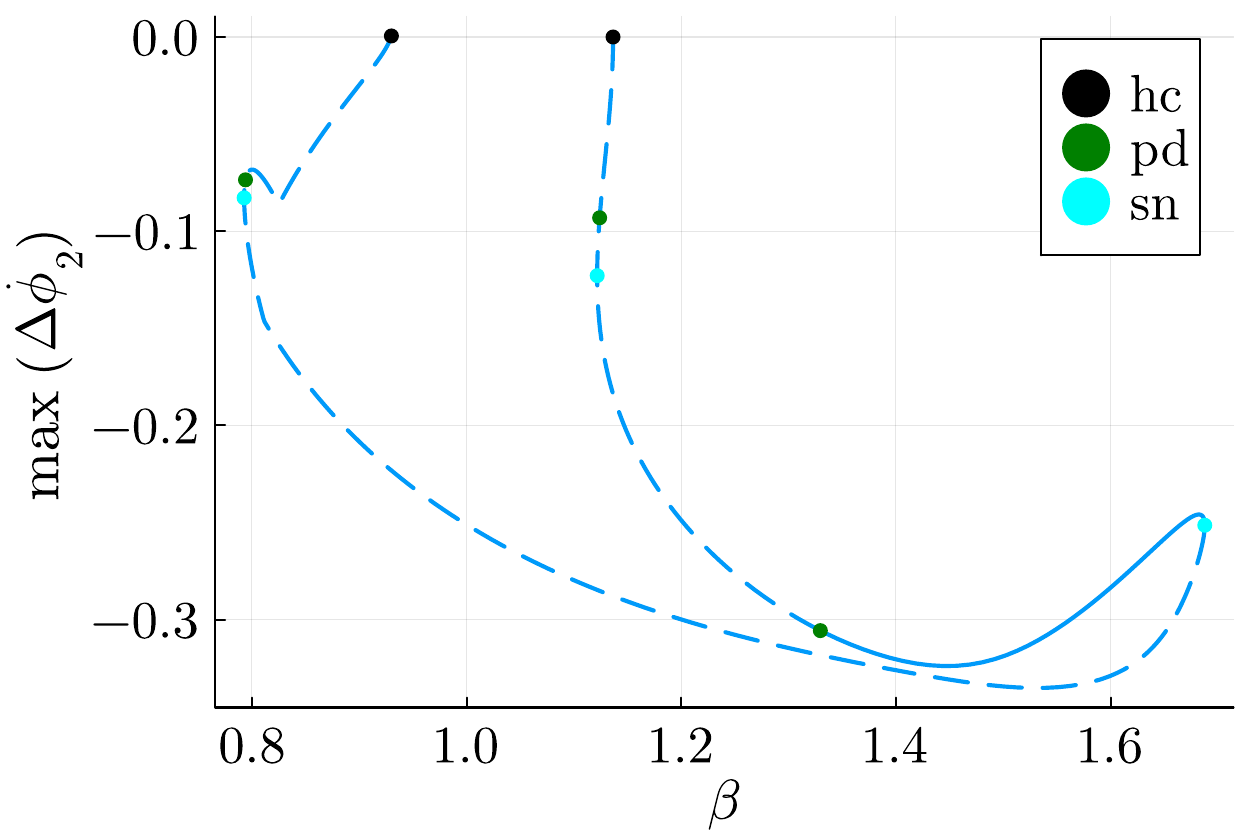}
	\caption[Continuation of the 4-2-1 state in the 3-cluster subspace]{\justifying Continuation of the 4-2-1 frequency-cluster state in the 3-cluster subspace. The maximum of the phase difference velocity $\Delta \dot{\phi}_2$ over one cycle is shown for varying $\beta$. Solid lines represent stable states and dashed lines depict unstable states. The colored dots indicate bifurcation points. The abbreviation hc denotes a homoclinic bifurcation, pd a period-doubling bifurcation, and sn a saddle-node bifurcation of limit cycles.}
	\label{fig:7_osc_subspace_4_2_1_state_continuation}
\end{figure}
\par
\begin{figure}[!h]
	\centering
	\includegraphics[width=0.45\textwidth]{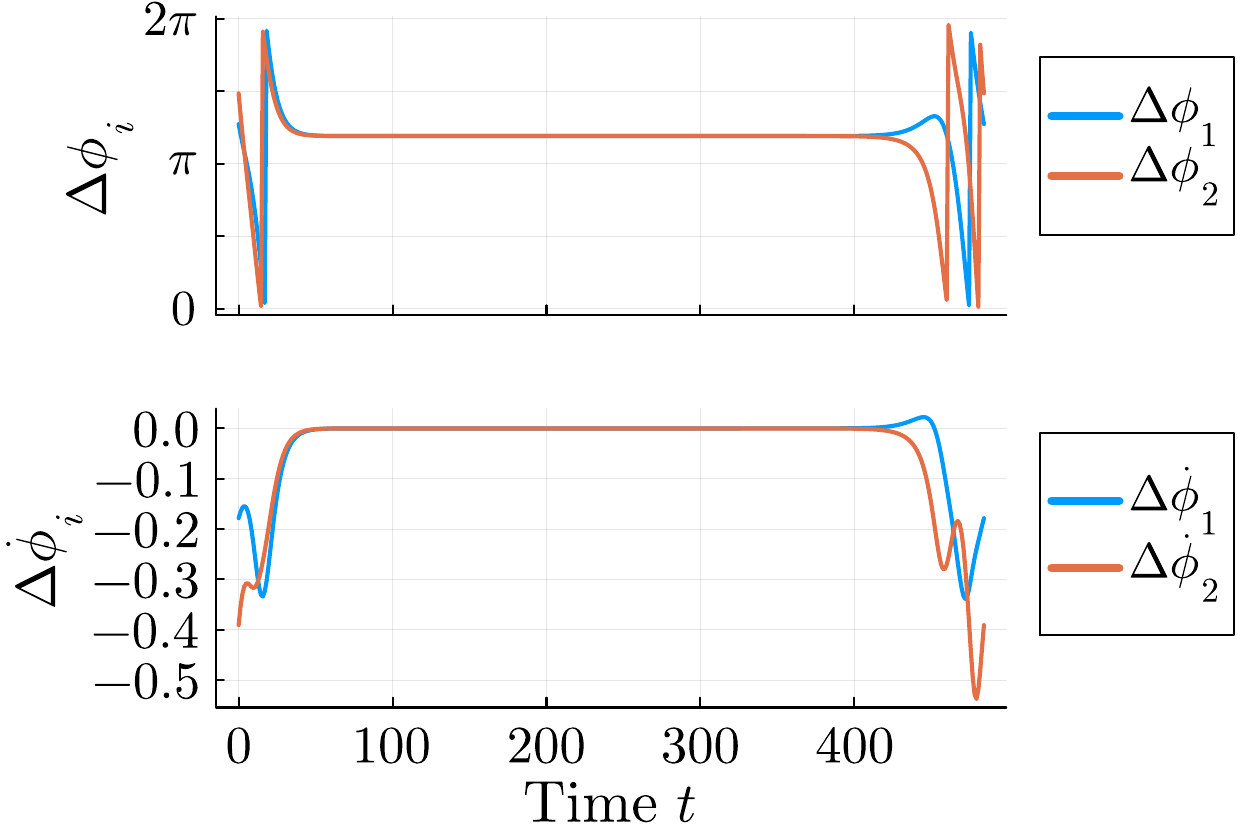}
	\caption[Homoclinic orbit of the 4-2-1 frequency-cluster state within the 3-cluster subspace at $\beta\approx 1.1365$]{\justifying Nearly homoclinic orbit for $\beta\approx 1.1365$ at one end of the continuation of the 4-2-1 frequency-cluster state within the 3-cluster subspace (cf. figure \ref{fig:7_osc_subspace_4_2_1_state_continuation}).}
	\label{fig:7_osc_subspace_continuation_homocline_time_series}
\end{figure}
\begin{figure*}
    \centering
  \begin{minipage}{0.375\textwidth}
    \centering
    \begin{overpic}[width=\linewidth]{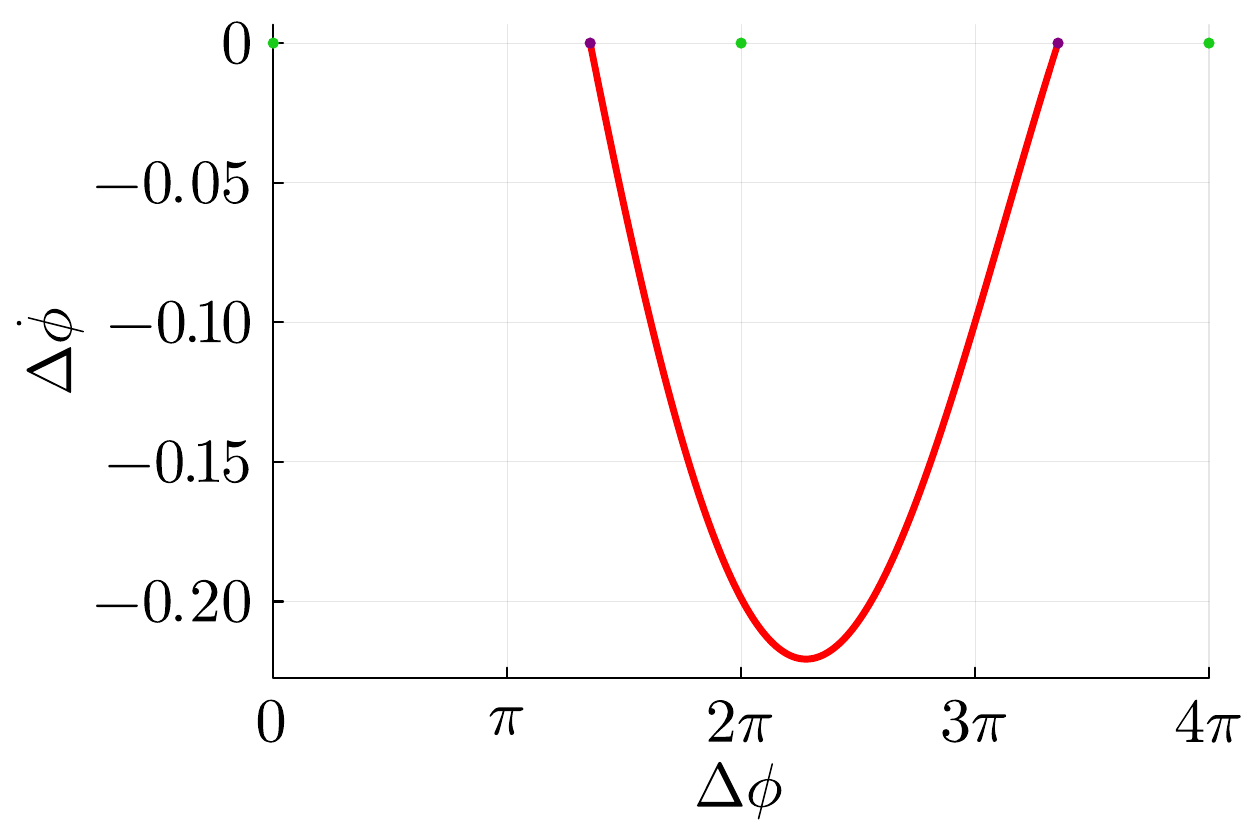}
        \put(10,70){(a)} 
    \end{overpic}
  \end{minipage}
  \hfill
  \begin{minipage}{0.58\textwidth}
    \centering
    \begin{overpic}[width=\linewidth]{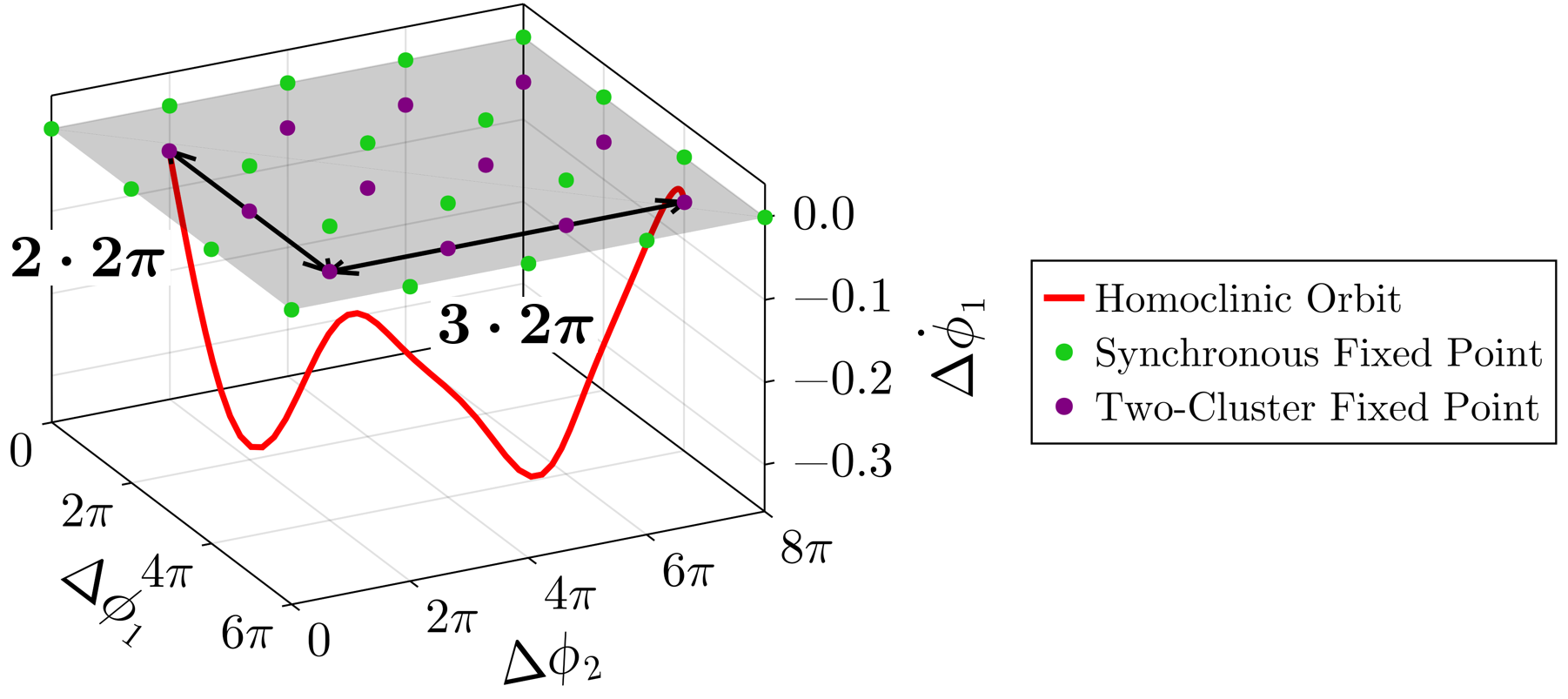}
        \put(5,46){(b)}
    \end{overpic}
  \end{minipage}
  \caption{\justifying Phase space illustration of homoclinic orbits, shown in red, within the cluster subspaces for $\rho_1=\frac{4}{7}$.
  The synchronous and the two-cluster fixed points, as well as their shifts in the phase differences by $2\pi$, are represented with colored dots. (a) displays a homoclinic orbit within the 2-cluster subspace from equations~\eqref{eq:KMI_2_cluster_phase_difference_equations}, that undergoes a shift of $2\pi$ in $\Delta \phi$. This orbit is involved in the bifurcation that creates two frequency clusters at $\beta=1.34543$. (b) shows a projection of the homoclinic orbit from figure~\ref{fig:7_osc_subspace_continuation_homocline_time_series} in phase space, where $\Delta \phi_1$ undergoes a shift of $2 \cdot 2\pi$ and $\Delta \phi_2$ a shift of $3 \cdot 2\pi$. The fixed points are all located within the shaded $\Delta\phi_1$-$\Delta\phi_2$-plane.
  The homoclinic orbit is involved in the bifurcation that creates three frequency clusters within the 3-cluster subspace for $\beta \approx 1.1365$ and $\rho_2=\frac{2}{3}$.}
  \label{fig:Homoclinic_orbits_2_and_3_cluster}
\end{figure*}

In the following paragraphs, we will examine the previously mentioned bifurcations in more detail. We start with the homoclinic bifurcation at $\beta \approx 1.1365$. In figure~\ref{fig:7_osc_subspace_continuation_homocline_time_series} a periodic solution close to the homoclinic connection is shown. During most of the time span, the state remains close to $\Delta \dot{\phi}_1 =\Delta \dot{\phi}_2=0$ and $\Delta \phi_1 = \Delta \phi_2 \approx 3.7391$. These coordinates correspond exactly to the two-cluster fixed point within the 2-cluster subspace from equation~\eqref{eq:fixed_point_two_clusters_phase_difference}, where $\rho_1=\frac{4}{7}$. Since the 2-cluster subspace is an invariant subspace of the 3-cluster subspace, all fixed points of the 2-cluster subspace are also contained within the 3-cluster subspace with the same value of $\rho_1$. This means that the 3-cluster subspace has a two-cluster fixed point, and this fixed point is involved in the homoclinic bifurcation that creates three frequency clusters. 
This fixed point is also involved in the other homoclinic bifurcation, which occurs at $\beta \approx 0.9300$. Therefore, the two bifurcations are qualitatively equivalent, and we do not discuss the one at $\beta \approx 0.9300$ further. 
Looking again at the evolution of the phase differences for the nearly homoclinic orbit in figure~\ref{fig:7_osc_subspace_continuation_homocline_time_series} , it is clear that the phase variables undergo multiple shifts of $2\pi$, as we already stated before. This is a necessary consequence of the implemented periodicity condition of the numerical continuation. 
However, we can also look at this from the opposite perspective. If a homoclinic bifurcation creates a frequency cluster, the homoclinic connection must emerge between a fixed point and its shift by multiples of $2\pi$ in at least one of its phase difference variables. If this shift does not occur, the emerging limit cycle would not exhibit a diverging, or rotating, phase difference, but only an oscillating or librating one, and the average frequency differences would be zero. The limit cycle would thus not represent a frequency cluster. \\
We illustrate these thoughts by plotting homoclinic orbits within phase space in figure~\ref{fig:Homoclinic_orbits_2_and_3_cluster}. In figure~\ref{fig:Homoclinic_orbits_2_and_3_cluster}(a), we show a homoclinic orbit within the 2-cluster subspace occurring at the homoclinic bifurcation that creates two frequency clusters, which we investigated in the previous section. It is visible that the homoclinic orbit shown in red connects two-cluster fixed points that are shifted by $2\pi$. Thereby, one rotation frequency with one diverging phase difference variable is created, resulting in two frequency clusters. In figure~\ref{fig:Homoclinic_orbits_2_and_3_cluster}(b), we show the homoclinic orbit from figure~\ref{fig:7_osc_subspace_continuation_homocline_time_series} in phase space, and project it into the 3-dimensional space spanned by the variables $\Delta\phi_1$, $\Delta\phi_2$ and $\Delta\dot{\phi}_1$.
Furthermore, the synchronous fixed point and the two-cluster fixed point, that is given by $\Delta\phi_1=\Delta\phi_2=2 \arctan \left( \frac{\cot(\beta)}{1-2\rho_1} \right)$ and $\Delta\dot{\phi}_1=\Delta\dot{\phi}_2=0$, and their shifts by $2\pi$ in both $\Delta\phi_1$ and $\Delta\phi_2$ are depicted. The homoclinic orbit, which is shown in red, connects two-cluster fixed points which are shifted by $2\cdot2\pi$ in $\Delta\phi_1$-direction and by $3\cdot2\pi$ in $\Delta\phi_2$-direction. Since the shifts in both variables are different, this creates a limit cycle with different rotation frequencies in $\Delta\phi_1$ and $\Delta\phi_2$.
Thus, we see that having shifts by different multiples of $2\pi$ is a prerequisite for obtaining two distinct mean frequency differences and thereby three frequency clusters. As we have shown before, it follows from these conditions that the state necessarily exhibits triplet locking. Therefore, a homoclinic bifurcation creating three frequency clusters always results in a triplet-locked state. \\
We can generalize these thoughts to systems with an arbitrary number of frequency clusters. If states with more than three frequency clusters are also created by homoclinic bifurcations, we would observe a pairwise locking of the mean frequency differences described by equation~\eqref{eq:locking_of_phase_differences}. Therefore, all possible combinations of three mean frequencies belonging to different clusters fulfill the condition for triplet-locking from equation~\eqref{eq:triplet_locking_condition}.

\begin{figure}[h!]
	\centering
	\includegraphics[width=0.45\textwidth]{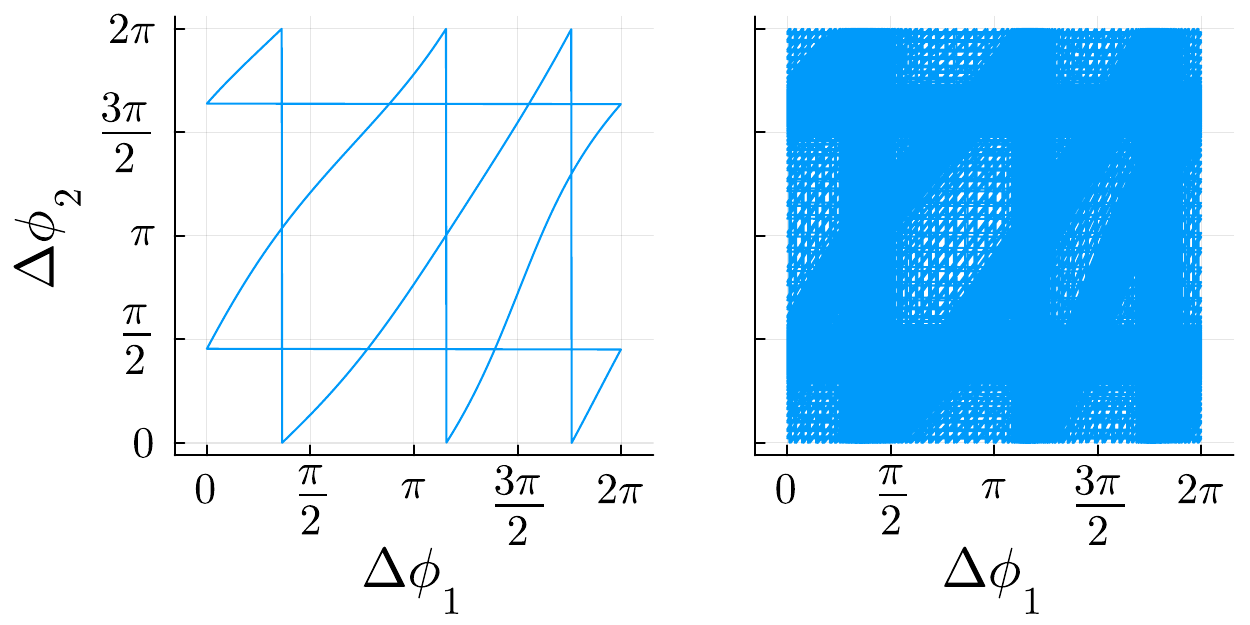}
	\caption[Unwrapped phase difference torus showing quasiperiodic behavior after the saddle-node bifurcation of the 4-2-1 frequency-cluster state in the 3-cluster subspace]{\justifying Phase difference dynamics on the unwrapped tori illustrating the saddle-node bifurcation of limit cycles of the 4-2-1 frequency-cluster state in the 3-cluster subspace. The left plot shows the state at the bifurcation point where $\beta \approx 1.6877$, while the right plot displays the state after the bifurcation at $\beta \approx 1.6887$ for 10000 time units. In both plots, the phase difference $\Delta \phi_1$ is plotted against $\Delta \phi_2$.}
	\label{fig:7_osc_subspace_4_2_1_state_sn_tori}
\end{figure}

Next, we turn to the period-doubling bifurcation at $\beta \approx 1.3296$, through which the three frequency state loses stability (cf. figure~\ref{fig:7_osc_subspace_4_2_1_state_continuation}). In this bifurcation, a new stable period-doubled state emerges within the 3-cluster subspace. Hence, this state still consists of three phase-synchronous frequency clusters.\\
\indent On the other side of the interval, in which the state is stable, a saddle-node bifurcation of limit cycles occurs at $\beta\approx 1.6877$. Here, the stable branch and the unstable branch, both of which were originally created in homoclinic bifurcations, participate in this bifurcation. We further investigate which attractor the dynamics converge to after the bifurcation. Therefore, we slightly increase $\beta$ by $10^{-3}$ and initialize the system in a point of the cycle at the saddle-node bifurcation. Figure~\ref{fig:7_osc_subspace_4_2_1_state_sn_tori} shows the corresponding result of numerical integration at values of $\beta$ slightly before and after the saddle node bifurcation of limit cycles. 
For both cases, the evolution of the phase differences $\Delta \phi_1$ and $\Delta \phi_2$ is depicted on an unwrapped torus.
On the left plot, where $\beta$ is chosen (within the numerical error) at the bifurcation point, it is apparent that the phase differences wind around the torus two respectively three times before the trajectory repeats. On the left plot, which was obtained for a value of $\beta$ beyond the bifurcation, the trajectory, which is shown for 10000 time units, does not repeat. An extension of the time span results in a further filling of the unwrapped torus. This indicates that the orbit is not periodic anymore and the phase differences do not wind around the torus with a rational relation. In other words, the trajectory is quasiperiodic \cite{Jensen.1984, Strogatz.2015}. Therefore, also the mean frequency differences are incommensurable and do not have a rational ratio anymore. Nevertheless, the ratio between the mean frequency differences, averaged over a sufficiently long time interval, remains close to the rational ratio observed before the bifurcation. This is a common phenomenon near saddle-node bifurcations, and the state is often referred to as a ghost of the bifurcation \cite{Strogatz.2015}.\\

\begin{figure}[h!]
	\centering
	\includegraphics[width=0.45\textwidth]{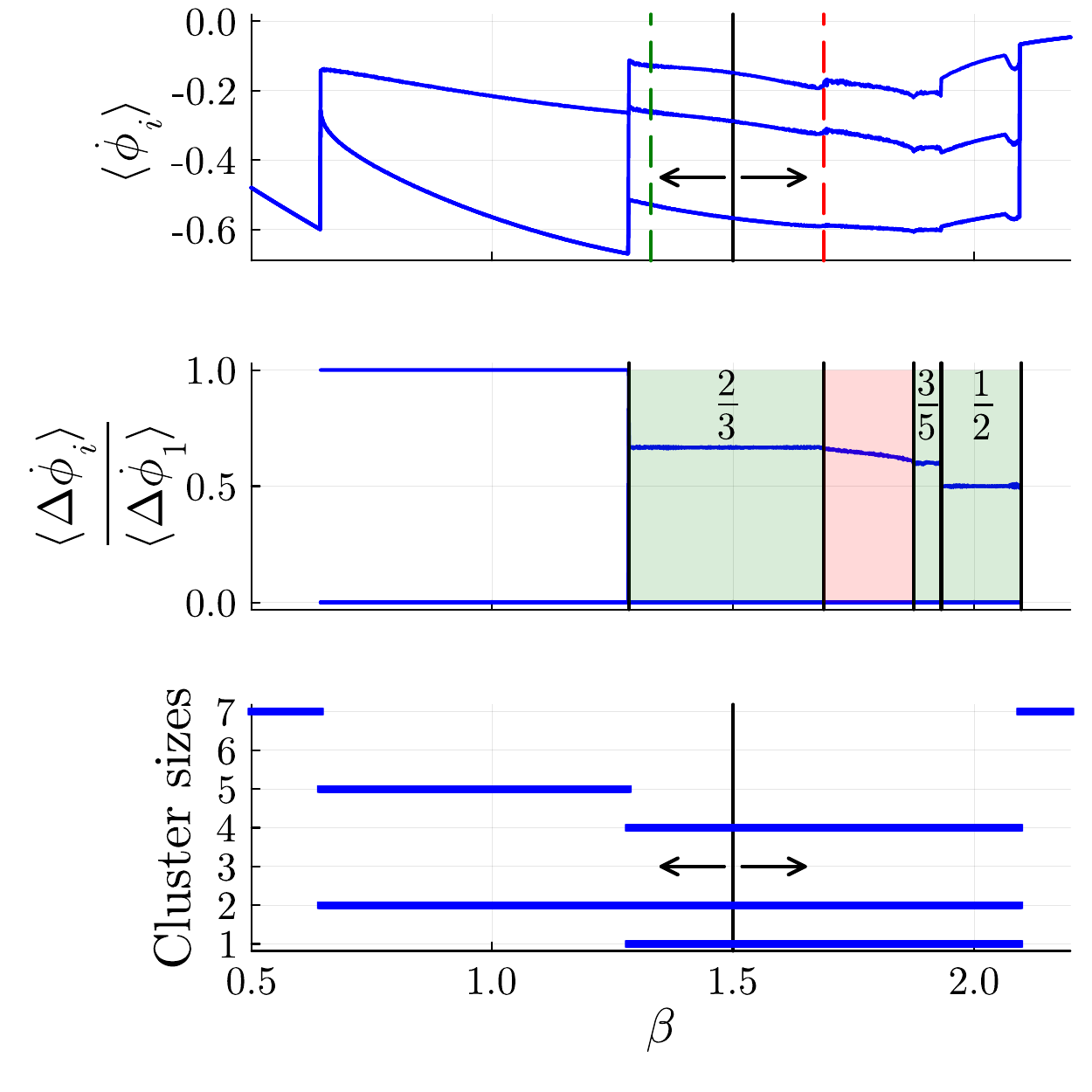}
	\caption[Parameter ramp for 7 oscillators without perturbation starting from the 4-2-1 state]{\justifying Parameter ramp without perturbation in a system of 7 oscillators, starting from a 4-2-1 state at $\beta=1.5$. The upper plot shows the mean frequencies $\langle \dot{\phi}_i \rangle$ averaged over 500 time units, the middle plot illustrates the ratios of the mean frequency differences, and the lower plot displays the cluster sizes as a function of $\beta$. The green and the red dashed lines
    mark the period-doubling and saddle-node bifurcations, respectively, in the continuation of the 4-2-1 state within the subspace (cf. figure~\ref{fig:7_osc_subspace_4_2_1_state_continuation}). The vertical lines with arrows indicate the initial value and the direction of change of $\beta$, which is varied in steps of $10^{-3}$. After each step, the system is integrated for a transient time of 10000 time units. The oscillator dynamics is governed by the \ac{kmi} from the equations~\eqref{eq:KMI_final_first_order_reduction}.
    }
	\label{fig:7_osc_parameter_scan_without_perturbation}
\end{figure}

In the following, we supplement the results from the numerical continuation with direct numerical simulations of the network with seven oscillators. Therefore, we conduct a parameter ramp, where we adiabatically change the parameter $\beta$ starting at the 4-2-1 state. It is essential that we do not introduce a perturbation to the state when changing $\beta$, ensuring that we never leave the 3-cluster subspace of the 4-2-1 state, which is guaranteed by the permutation symmetry. Once two oscillators have exactly equivalent state variables, even numerical errors do not split synchronous oscillators, since the errors are equal for both. The result of the parameter ramp is displayed in figure~\ref{fig:7_osc_parameter_scan_without_perturbation}. Here, the mean frequencies $\langle \dot{\phi}_i \rangle$, the ratios of the mean frequency differences $\langle \Delta\dot{\phi}_i \rangle$ ($i \in \{2,\ldots,6\}$) to the largest mean frequency difference $\langle \Delta\dot{\phi}_1\rangle$, and the sizes of clusters with equal mean frequency are shown as a function of $\beta$. Starting from the stable 4-2-1 state at $\beta=1.5$, indicated by the vertical line, the parameter $\beta$ is once decreased and once increased, as illustrated by the arrows. \\
As $\beta$ decreases, the 4-2-1 state remains stable until $\beta = 1.283$, where it undergoes a transition to a 5-2 two frequency state. 
A comparison with the longitudinal period-doubling bifurcation of the 4-2-1 state at $\beta \approx 1.3296$ indicated by the green dashed line and detected in the continuation from figure~\ref{fig:7_osc_subspace_4_2_1_state_continuation}, clearly shows that the two values differ. In the interval $\beta \in [1.283,1.3296)$, the parameter ramp presumably reaches period-doubled states and potentially further period-multiplied states.\\
For increasing values of $\beta$ starting from $\beta=1.5$, 4-2-1 states with three distinct frequency clusters remain stable until a value of $\beta$ above 2. A peculiarity in the ramp are the kinks in the mean frequencies. The first kink after the starting value $\beta=1.5$ appears at $\beta \approx 1.69$. This value coincides with the saddle-node bifurcation occurring at $\beta \approx 1.6877$, marked by the red dashed line. This kink most likely indicates the change from the periodic to the quasiperiodic regime that is shown in figure~\ref{fig:7_osc_subspace_4_2_1_state_sn_tori}. Before this kink, the mean frequency differences exhibit the rational ratio of $\frac{2}{3}$, and after the kink, the ratio seems to decrease continuously. 
In this region of decreasing ratio, we also solved the \acp{ode} for different initial conditions and found states with other ratios of mean frequency differences, such as $\frac{5}{8}$ or even $\frac{32}{51}$. Nevertheless, during this parameter ramp such states with different ratios were not observed. Instead, the trajectories remained quasiperiodic, which suggests that in this $\beta$ interval quasiperiodicity is co-stable with periodic states. However, it is also possible that the stability interval of these locked states is too narrow to be resolved in the parameter scan.
After the area where the frequency difference ratio decreases continuously, a small locked region with a ratio of $\frac{3}{5}$ exists, which coincides with further kinks in the mean frequencies. Further increasing $\beta$, the frequency differences undergo a transition to a $\frac{1}{2}$ locking. At some point, this 3-cluster limit cycle disappears and the system transitions towards a 1-cluster state.\\
In the region with constant ratios of frequency differences, we additionally show ratios of mean frequencies during the parameter ramp in figure \ref{fig:7_osc_parameter_scan_frequency_ratios}, where the vertical lines separate the same regions as in figure~\ref{fig:7_osc_parameter_scan_without_perturbation}. 
The mean frequency of an oscillator from the large cluster (size four) serves as the reference in the upper plot, whereas the medium cluster (size two) provides the reference for the lower plot. It is visible that in most regions, the ratios are not constant, meaning that the oscillators are not locked pairwise (except oscillators in the same cluster, corresponding to a ratio of one). Only in the narrowest region, where the state exhibits a $\frac{3}{5}$ locking in frequency difference this is not obvious. However, a thorough examination of this state would presumably show varying ratios.
The absence of pairwise locking together with the locking of mean frequency differences, described by equation~\eqref{eq:triplet_locking_condition}, imply that these states exhibit triplet locking according to the definition of Kralemann et al.\cite{Kralemann.2013}. The ratio $\frac{2}{3}$ corresponds to a $2:-3:1$ locking, $\frac{3}{5}$ to $3:-5:2$, and $\frac{1}{2}$ to a $1:-2:1$ locking.
\begin{figure}[h!]
	\centering
	\includegraphics[width=0.45\textwidth]{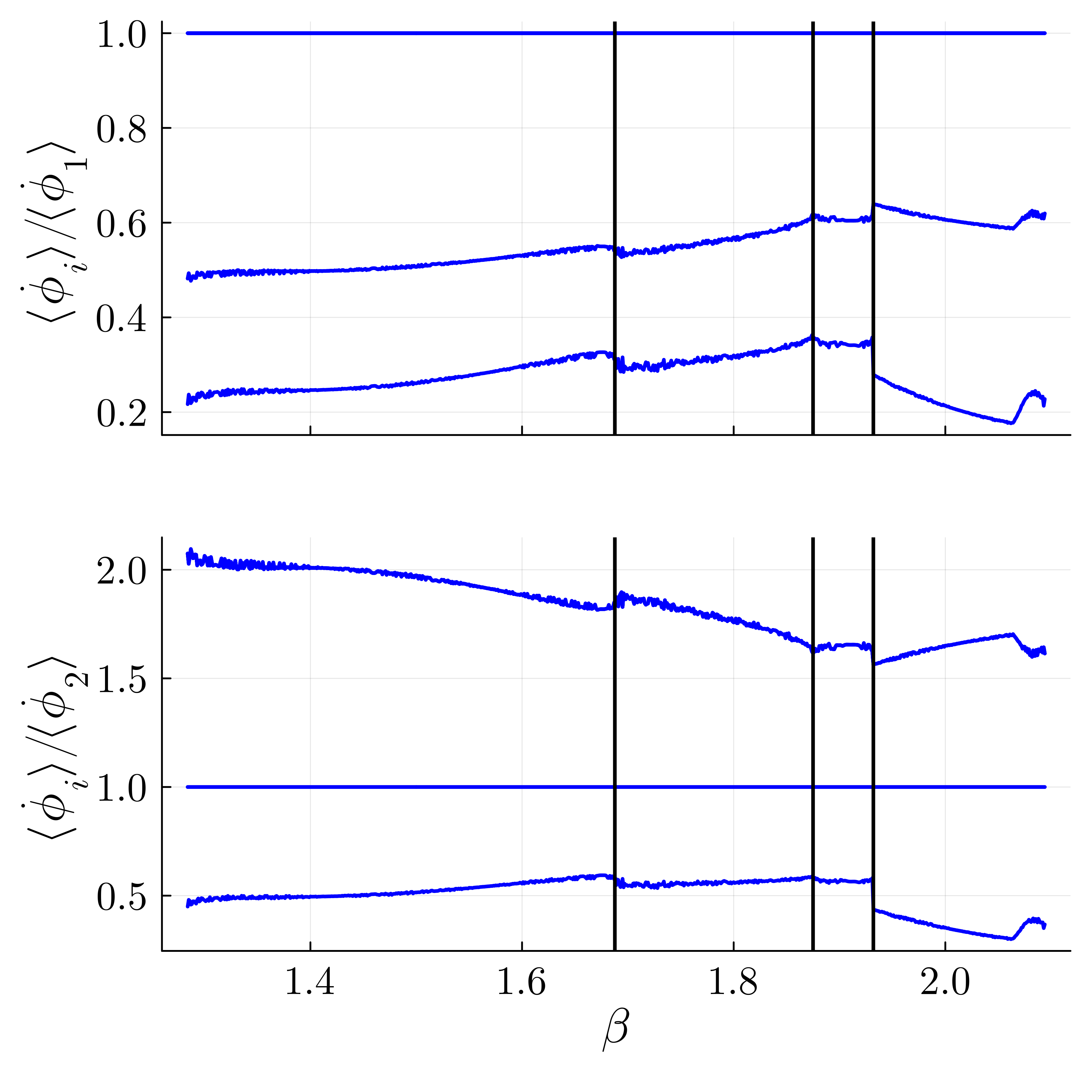}
	\caption{\justifying Ratios of mean frequencies 
    during the parameter ramp (cf. figure \ref{fig:7_osc_parameter_scan_without_perturbation}), with oscillators 1 and 2 residing in different clusters. No clear pairwise frequency locking between oscillators of different clusters is visible.
    }
	\label{fig:7_osc_parameter_scan_frequency_ratios}
\end{figure}

\par
\paragraph*{Arnol'd tongues}
The $\beta$ intervals, in which states with locked frequency differences were found, i.e., for which the phase differences wind around the torus with a rational relation, are reminiscent of Arnol'd tongues that denote parameter regions in the detuning - coupling strength parameter plane where forced or coupled oscillators exhibit frequency locking \cite{Jensen.1984}. Arnol'd tongues are bounded by saddle node bifurcations of limit cycles, outside of which quasiperiodic motion prevails. In our continuation, shown in  figure~\ref{fig:7_osc_subspace_4_2_1_state_continuation}, we also found that a saddle-node bifurcation at $\beta \approx 1.6877$ destroys the periodic state and the trajectory settles to a quasiperiodic motion. In addition, changing $\beta$, which implicitly changes the frequencies of the clusters, we observed a series of entrained regions characterized by a monotonically decreasing rational number. However, a notable difference is that for 'classical' Arnol'd tongues, the locking occurs between frequencies, not frequency differences as we observed it here. This points to the existence of a generalized mechanism that unifies classical 'doublet' locking of two frequencies and 'triplet' locking, i.e. the coexistence of three frequencies with rational frequency differences. 

\subsection{Transversal Stability of Three Frequency Clusters}
\begin{figure}[h!]
	\centering
	\includegraphics[width=0.45\textwidth]{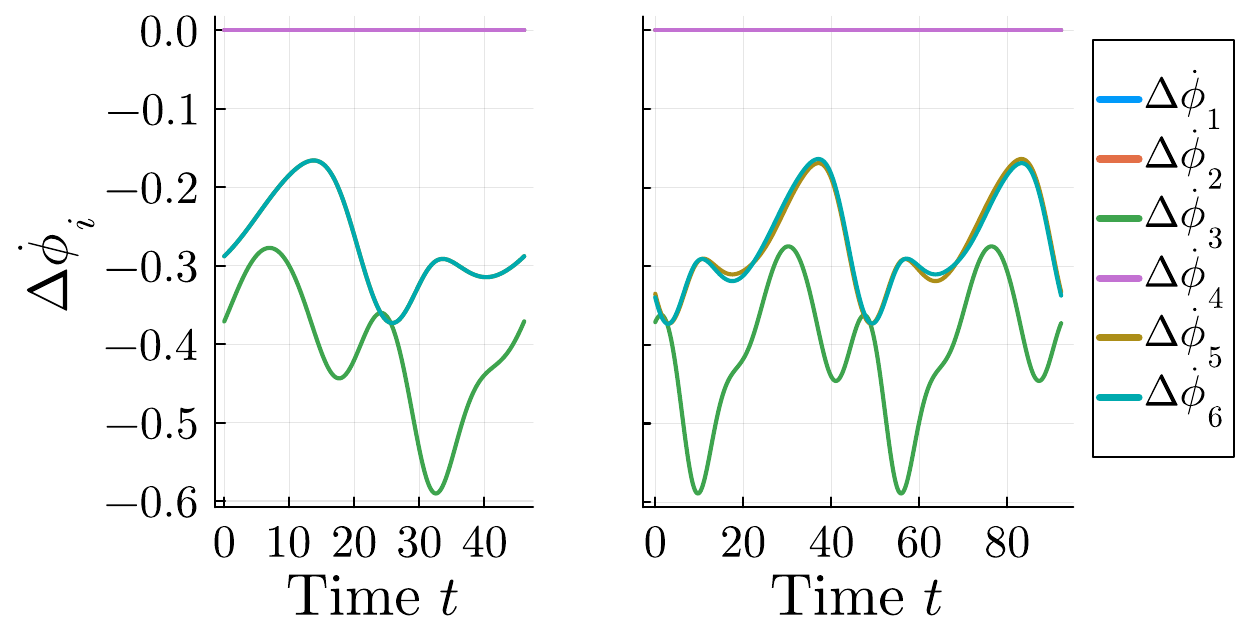}
	\caption[Period-doubling bifurcation of the 4-2-1 frequency-cluster state at $\beta \approx 1.6127$]{\justifying Transversal period-doubling bifurcation of the 4-2-1 frequency-cluster state. The plot on the left shows the evolution of the frequency differences $\Delta \dot{\phi}_i$ at the bifurcation point, where $\beta \approx 1.6127$, and the plot on the right displays the frequency differences $\Delta \dot{\phi}_i$ after the bifurcation at $\beta \approx 1.6227$.}
	\label{fig:7_osc_4_2_1_state_period_doubling_right}
\end{figure}

After studying the longitudinal stability of three frequency clusters in the previous part, we discuss their transversal stability in this section. We do this by continuing the 4-2-1 state from figure~\ref{fig:7_osc_3_clusters_time_series} in the full system with seven oscillators described by equations~\eqref{eq:KMI_final_phase_difference_coordinates}. Without the restriction to the 3-cluster subspace from the previous section, transversal bifurcations that split up the clusters are detected as well. The continuation is displayed in figure~\ref{fig:7_osc_4_2_1_state_continuation}. Comparing it with the continuation within the 3-cluster subspace from figure~\ref{fig:7_osc_subspace_4_2_1_state_continuation}, it is visible that additional bifurcations occur, which are transversal bifurcations. The frequency clusters are stable for $\beta \in [1.3289, 1.6127]$ and at both ends of this interval, the state is destabilized by period-doubling bifurcations. At the lower end of the interval, this is the longitudinal period-doubling bifurcation described above. 
At the upper end, a transversal period-doubling bifurcation occurs. We visualize this bifurcation in figure~\ref{fig:7_osc_4_2_1_state_period_doubling_right}, where the evolution of the frequency differences $\Delta \dot{\phi}_i$ at the bifurcation is shown on the left, and the stable state after the bifurcation on the right. 
It is visible that after the bifurcation, the time series of the frequency differences $\Delta \dot{\phi}_5$ and $\Delta \dot{\phi}_6$ are not equivalent anymore. Hence, the cluster consisting of 2 oscillators loses its synchrony through this bifurcation but still oscillates with the same mean frequency. Thus, this period-doubling bifurcation creates a frequency-cluster state with one phase-asynchronous cluster.
\begin{figure}[!h]
	\centering
	\includegraphics[width=0.45\textwidth]{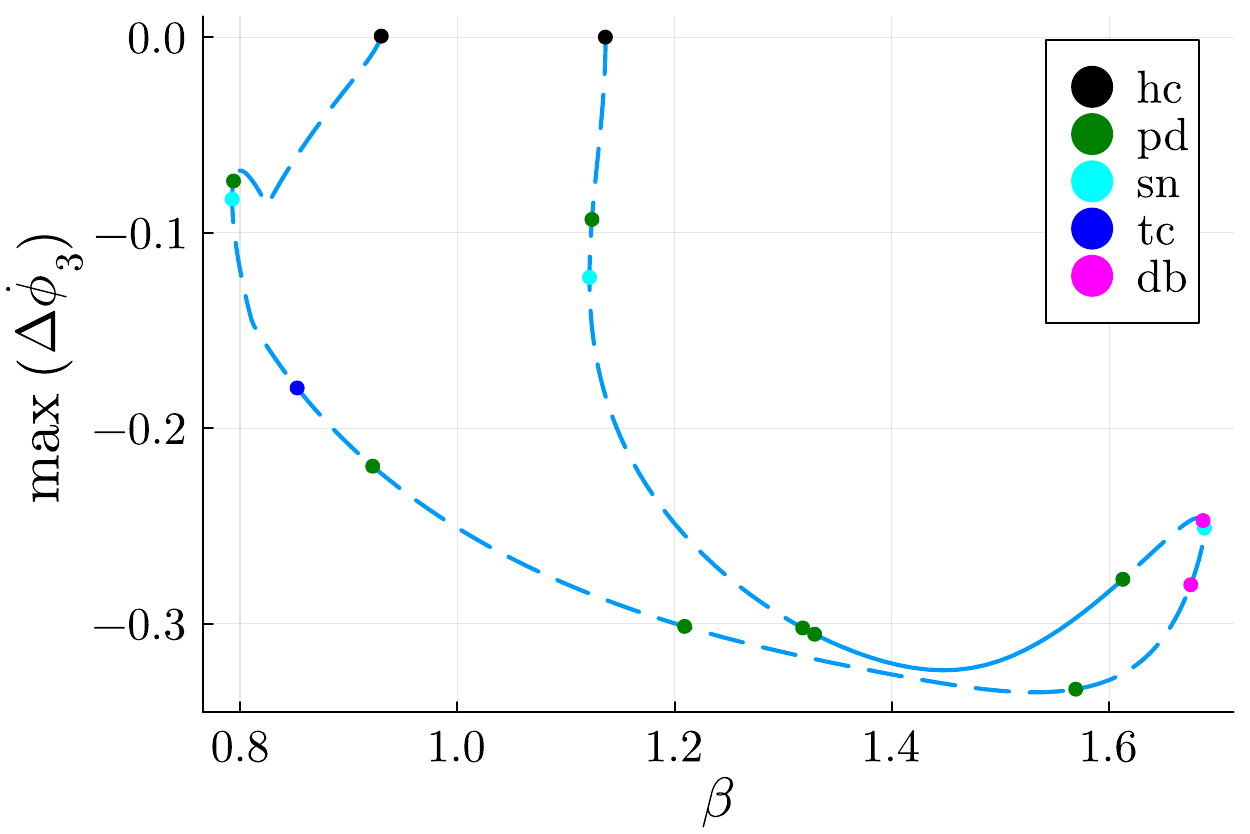}
	\caption[Continuation of the 4-2-1 frequency-cluster state]{\justifying Continuation of the 4-2-1 frequency-cluster state. The maximum of the phase difference velocity $\Delta \dot{\phi}_3$ over one cycle is shown for varying $\beta$. Solid lines represent stable states and dashed lines depict unstable states. The colored dots indicate bifurcation points, with relevant ones described in the text. The abbreviation hc denotes a homoclinic bifurcation, pd a period-doubling bifurcation, sn a saddle-node bifurcation of limit cycles, tc a transcritical bifurcation point where a single real Floquet multiplier crosses +1, and db a degenerate bifurcation involving multiple real Floquet multipliers crossing +1.}
	\label{fig:7_osc_4_2_1_state_continuation}
\end{figure}

\section{Impossibility of Hopf Bifurcations Creating Frequency Clusters in Phase Oscillators}
\label{sec:Impossibility_of_a_Hopf_Bifurcation_to_Create_Frequency_Clusters}
As we have seen, the bifurcation creating two and three frequency clusters in the \ac{kmi} is the homoclinic bifurcation. 
Of course, this might not be the only possible bifurcation creating frequency clusters. One of the bifurcations that often occurs when an oscillation emerges is the Hopf-Andronov bifurcation, which we will refer to as Hopf bifurcation in the following. Therefore, one may wonder whether this bifurcation can possibly create frequency clusters. In the vicinity of a Hopf bifurcation, the amplitude of the limit cycle in phase space scales with $\mathcal{O}(\sqrt{\mu}) \ $ \cite{Strogatz.2015}, assuming the bifurcation occurs at $\mu=0$ and the limit cycle emerges for $\mu>0$. Thus, the limit cycle remains confined to a small and bounded region of phase space and shrinks to a single point in all variables for $\mu \rightarrow 0$. However, a frequency-cluster state exhibiting different mean frequencies, corresponding to a non-zero mean frequency difference $\langle \Delta\dot{\phi} \rangle$, is characterized by at least one phase difference $\Delta\phi$ that diverges. 
Therefore, the cycle can never be confined to an arbitrarily small and bounded region with a single point in its interior. Consequently, for phase oscillators, it is impossible for frequency clusters to be created by a Hopf bifurcation. 
This can be illustrated with figure~\ref{fig:phase_space_2_cluster_subspace}.
If a Hopf bifurcation occurred, a cycle around one of the fixed points would emerge. 
This would create a libration, but never a rotating limit cycle highlighted by the red curve which is a characteristic of frequency clusters.

To formalize this argument, we first recall the definition of a bifurcation as "the appearance of a topologically nonequivalent phase portrait" \cite{Leine.2000}. Consequently, we need to show that a limit cycle representing frequency clusters is topologically different to a limit cycle created by a Hopf bifurcation. This would imply that another bifurcation must occur after the Hopf bifurcation to create a cycle that can represent frequency clusters.\\
A limit cycle created by a Hopf bifurcation is confined to a bounded interval in the phase (difference) variable, respectively when considering all variables it lies inside a bounded $n$-dimensional ball $B^n$ due to the scaling behavior of the Hopf bifurcation. The topology of these spaces is characterized by the fundamental group $\pi_1$.
The fundamental group $\pi_{1}(X)$ of a space $X$ is a topological invariant describing the structure of loops in the space up to continuous deformation \cite{Hatcher.2018}. We write $\pi_{1}(X) \cong G$ when $\pi_{1}(X)$ is isomorphic to a group $G$, meaning that both groups share the same algebraic structure.
For the space in which the cycle created by a Hopf bifurcation is located, we have that $\pi_1(B^n) \cong \pi_1(\mathbb{R}) \cong \{0\}$ \cite{Hatcher.2018}. In contrast, a limit cycle describing frequency clusters requires at least one phase difference to be defined on the circle $S^1$, such that the whole cycle would be embedded in a space of the form $S^1 \times \mathbb{R}^{n-1}$. Their fundamental group is given by $\pi_1 (S^1 \times \mathbb{R}^{n-1}) \cong \pi_1 (S^1) \cong \mathbb{Z}$ \cite{Hatcher.2018}. Since the two limit cycles reside in state spaces with non-isomorphic fundamental groups, their topology is distinct. It follows that for phase oscillators, a limit cycle exhibiting frequency clustering cannot emerge directly from a Hopf bifurcation.\\
The argument about the topology of the space in which the limit cycle resides holds only for local bifurcations, such as the Hopf bifurcation, but not for other global bifurcations. 
For example, we found that two states representing three frequency clusters can merge and disappear in a saddle-node bifurcation of limit cycles, which is therefore a possible bifurcation that can be involved in the creation of frequency clusters. 
Other global bifurcations that might create frequency clusters are heteroclinic bifurcations or saddle-node bifurcations of infinite period.

\section{Conclusion and Outlook\label{sec:C&O}}
In this work, we investigated the creation of two and three frequency clusters in the \ac{kmi}. We examined two frequency clusters in the thermodynamic limit and confirmed that their origin are homoclinic bifurcations, as reported by Belykh et al. \cite{Belykh.2016}. We extended previous works \cite{Munyayev.2024, Ashwin.2025} by analyzing the transversal stability of the two clusters via a numerical bifurcation analysis, using the 3-cluster subspace to assess the stability via a test cluster with zero weight. We found that the phase synchrony of both clusters is destroyed via transverse transcritical or transverse period-doubling bifurcations. 
In the continuations of these transversal bifurcations in the parameters $\beta$ and $\rho_1$, codimension-2 points involving branches of transcritical, period-doubling, and homoclinic bifurcations emerge. We identify one of the codimension-2 points as the analogue of a point found by Ashwin and Bick \cite{Ashwin.2025} in a system of three oscillators. Furthermore, when analyzing the transversal stability of the large cluster, we find a sequence of such codimension-2 points which is reminiscent of a scenario recently reported for a network of mean-field coupled Stuart-Landau oscillators \cite{Thome.2025}.
As the dynamics of the latter system differ significantly from those of the KMI, these codimension-2 points might be a characteristic motif organizing the destabilization of oscillatory cluster solutions in systems of full permutation symmetry.\\
Next, we studied the origin of three frequency clusters in a network of seven oscillators. We found that three synchronous frequency clusters can be destabilized by longitudinal and transversal period-doubling bifurcations, and that three frequency clusters also originate from homoclinic bifurcations. If frequency clusters are created by homoclinic bifurcations, the frequency differences necessarily exhibit a rational ratio due to the integer shift of multiples of $2\pi$ in the phase differences within the homoclinic orbit. For three or more frequency clusters, this also implies a triplet locking, as defined by Kralemann et al. \cite{Kralemann.2013}. Furthermore, we found that irrational ratios of frequency differences can appear when the state with a rational ratio is destroyed by a saddle-node bifurcation of limit cycles. 
These features strongly resemble those of Arnol'd tongues. Finally, we made some general considerations about the creation of frequency clusters. We demonstrated that frequency clusters in phase oscillators cannot be created by a Hopf bifurcation, elucidating the importance of global bifurcations for the occurrence of frequency clusters. \par
While we studied two and three frequency clusters and uncovered mechanisms governing their formation and destabilization, this work raises several further questions.
At first, the emergence of the sequence of codimension-2 points that organize the transversal bifurcation of two clusters requires further investigation. It will remain a challenge to extend the results of Ashwin and Bick \cite{Ashwin.2025} about a single codimension-2 point in a system of three oscillators to the sequence of such points existing in large oscillator systems and in the thermodynamic limit. Here, it is also unclear why this sequence only occurs for the transversal bifurcations of the large cluster and not for those of the small cluster.
Second, the relationship between the locking behavior we observed and the frequency locking known from Arnol'd tongues is an interesting topic for future work. The main difficulties here are the second-order nature of the \acp{ode} and the dependence of the frequency difference on, especially, the cluster size and $\beta$. 
Finally, an open question concerns the origin of cluster states with more than three frequencies. In particular, four frequency clusters have already been observed in a system with 100 oscillators \cite{Berner.2021}. 
Detecting bifurcations to these states poses significant numerical challenges. However, given the potential significance of frequency clusters in the context of adaptive networks and their functionalities, it seems to be important to address these challenges in the future. 


\begin{acknowledgments}
The authors thank Christian Bick for help with the AUTO-07p scripts.
\end{acknowledgments}

\section*{Abbreviations}
\begin{acronym}
    \setlength{\itemsep}{-0.5cm}
    \acro{ode}[ODE]{ordinary differential equation}
    \acro{kmi}[KMI]{Kuramoto model with inertia}
\end{acronym}

\section*{References}
\bibliography{Literature_master_thesis}

\end{document}